\documentclass[12pt, draftclsnofoot, onecolumn]{IEEEtran}
\usepackage{geometry}
\ifCLASSINFOpdf

\else

\fi
\usepackage[cmex10]{amsmath}
\usepackage{amssymb}
\usepackage{comment}
\usepackage{cite}
\usepackage{graphicx}
\usepackage{array,color}
\usepackage{amsmath}
\usepackage{stfloats}
\usepackage{graphicx}
\usepackage{subfigure}
\usepackage{tabularx}
\usepackage{epsfig,epsf,color,balance,cite}
\usepackage{algorithmic}
\usepackage{algorithm}
\usepackage{bm}
\usepackage{textcomp}

\begin{document}

\title{ Intelligent Reflecting Surface-aided URLLC in a Factory Automation Scenario}
\author{Hong Ren, Kezhi Wang and  Cunhua Pan
\thanks{ H. Ren is with National Mobile Communications Research Laboratory, Southeast University, Nanjing, China. (e-mail:hren@seu.edu.cn).  K. Wang is with Department of Computer and Information Sciences, Northumbria University, Newcastle, UK. (e-mail: kezhi.wang@northumbria.ac.uk). C. Pan is with the School of Electronic Engineering and Computer Science at  Queen Mary University of London, U.K. (e-mail:c.pan@qmul.ac.uk).}}

\maketitle

\begin{abstract}
Different from conventional wired line connections, industrial control  through wireless transmission is widely regarded as a promising solution due to its reduced cost, increased long-term reliability, and enhanced reliability. However, mission-critical applications impose stringent quality of service (QoS) requirements that entail ultra-reliability low-latency communications (URLLC). The primary feature of URLLC is that the blocklength of channel codes is short, and the conventional Shannon's Capacity is not applicable. In this paper, we consider the URLLC in a factory automation (FA) scenario. Due to  densely deployed equipment in FA, wireless signal are easily blocked by the obstacles. To address this issue, we propose to deploy intelligent reflecting surface (IRS) to create an alternative transmission link, which can enhance the transmission reliability. In this paper, we focus on the performance analysis for IRS-aided URLLC-enabled communications in a FA scenario. Both the average data rate (ADR) and the average decoding error probability (ADEP) are derived under finite channel blocklength for seven cases: 1) Rayleigh fading channel; 2) With direct channel link; 3)  Nakagami-m fading channel; 4) Imperfect phase alignment; 5) Multiple-IRS case; 6) Rician fading channel; 7) Correlated channels. Extensive numerical results are provided to verify the accuracy of our derived results.
\end{abstract}


\IEEEpeerreviewmaketitle
\section{Introduction}
Nowadays, industry world is evolving into another new paradigm, that is the so-called fourth industrial revolution or Industrial 4.0 \cite{schwab2017fourth}, in which more advanced manufacturing functions can be realized with the aid of industrial internet-of-things (IIoT).
Conventionally, industrial control systems mainly rely on wired network infrastructure, where the central control unit is connected to various machines through wired lines such as copper lines or optical fibers. However, there are some drawbacks by deploying wired lines such as   high installation and maintenance costs, limited motion ranges, and vulnerability to wear and tear in the applications with motion actions.

As a result, the adoption of wireless networks is a promising solution to bypass all the aforementioned issues associated with wired lines. However, the industrial communications are totally different from conventional wireless communications. Instead, they require deterministic communications with the stringent quality of service requirements such as ultra reliability and low latency communications by exchanging a small amount of data such as control command data or measurement data. For typical industrial scenarios such as factory automation (FA), the maximum transmission latency is required to be kept within one millisecond, and the packet error probability should range from $10^{-6}$-$10^{-9}$ \cite{zhibo2017}.  In addition, due to the densely deployed equipment such as metal machinery, random movement of objects (robots and trucks), and thick pillars,  the wireless signals are more vulnerable to blockages, which can degrade the reliability performance.

To overcome the above hurdle, intelligent reflecting surface (IRS) is regarded as a promising solution, which has attracted considerable research interest from both academia and industry \cite{wu2019towards,Basar2019,pan2020reconfigurable}. In particular, IRS is a square or circular panel that consists of a large number of passive reflecting elements, each of which can induce an independent phase shift on the impinging electromagnetic (EM) waves. Hence, by carefully designing the phase shifts of all reflecting elements, the reflected EM waves can be constructively added with the direct signal from the base station (BS) to enhance the signal power at the desired user, which can enhance the system signal-to-noise (SNR) performance.

Most of the existing contributions on IRS focus on the transmission design for various IRS-aided wireless networks by jointly optimizing the active beamforming at the BS and the phase shift matrix at the IRS for perfect channel state information (CSI) in \cite{xianghaoyu2019,shuowen,qingqingwutwc,boyadi2020,cunhua202twc,pan2020jsac,zhengchu2020,huiming2020,hong2020perfectcsi,tong2020}, imperfect CSI in  \cite{guizhouwcl,xianghaoyu2020,guirobusttsp,hong2020robust}, or statistical CSI in \cite{yuhan2019,zhangjie,kangda,liyou2021}.
Although extensive research efforts have been devoted to the transmission design for IRS-aided networks, only a few works studied the analytical performance in IRS-aided wireless systems \cite{emil2020,Boulogeorgos2020,macroojcs,Basar2019,liangyang2020,liang2020,qintao2020,Ibrahim2021,Atapattu2020,jiayi2021,Badiu2020,Javier2020,pengxu2020,dongli2020}.
The performance comparison between IRS-aided systems and relaying systems was performed in \cite{emil2020,Boulogeorgos2020,macroojcs}. For the IRS-aided single  user communication system, the symbol error probability (SEP) was derived for the Rayleigh channel in \cite{Basar2019}, and the coverage probability was analyzed in  \cite{liangyang2020}. The central limit theorem (CLT) was used in \cite{Basar2019} and \cite{liangyang2020} to derive the probability density function (PDF) of signal-to-noise ratio (SNR). Then the Rayleigh channel model was extended to the more general Rician channel in \cite{liang2020,qintao2020} and Nakagami channel in \cite{Ibrahim2021}. The outage probability and spectral efficiency were analyzed in \cite{Atapattu2020} for IRS-aided two-way networks. The exact distribution for signal-to-noise ratio (SNR) was derived in \cite{jiayi2021} for an IRS-aided millimeter wave (mmWave) communication system, where the fluctuating
two-ray   distribution was adopted to characterize the small-scale fading in the mmWave band. In practical systems, the phase shifts can only be set with finite discrete values, which induce phase errors. Hence, recent contributions have focused on the study of the impact of the phase errors on the system performance \cite{Badiu2020,Javier2020,pengxu2020,dongli2020}. The authors in \cite{Badiu2020} demonstrated that transmission under imperfect reflectors is equivalent to a point-to-point communication over a Nakagami  fading channel. For the IRS-aided physical layer security  networks with phase errors, secrecy outage probability and the average secrecy rate was studied in \cite{Javier2020} for a single eavesdropper case and  in \cite{pengxu2020} for the case with multiple eavesdroppers.

All the above-mentioned contributions only focused on the traditional services with long packet, in which Shannon's Capacity is an accurate approximation of the achievable data rate. However, in a factory automation scenario with URLLC requirements, only small packets are transmitted due to the delay limit. In this case, the channel blocklength is finite, and the decoding error probability cannot approach zero for arbitrarily high SNR. Then, Shannon's Capacity can no longer be applied \cite{Durisi2016} since the law of large numbers does not hold and it cannot characterize the maximal achievable data rate with given decoding error probability. In fact, simulation results in \cite{schiessl2015delay} demonstrated that the delay outage probability will be underestimated if directly using Shannon's Capacity, and thus the quality of service (QoS) requirements cannot be guaranteed. In the seminal work in \cite{polyanskiy2010channel}, by using normal approximation technique  the authors have derived the accurate approximation of the maximal achievable rate that is valid under the short blocklength regime for the point-to-point AWGN channel, which is a complicated function of SNR, channel blocklength, and decoding error probability. Most recently, based on the results in \cite{polyanskiy2010channel}, extensive research attention has been devoted to the short packet transmission design \cite{xiaofang2018,Ghanem2019,SHe2108tcom,Avranas2018,jiechen2019,yulinhu2019,hongrentwc,hongrenjsac} or performance analysis under the short blocklength regime \cite{Zheng2019,Schiessl2018,jiezeng2020,hongren2019wcl}. However, all these contributions in \cite{xiaofang2018,Ghanem2019,SHe2108tcom,shuhan2019,Avranas2018,jiechen2019,yulinhu2019,hongrentwc,hongrenjsac,Zheng2019,Schiessl2018,jiezeng2020,hongren2019wcl}
did not consider the IRS-aided wireless systems. To the best of our knowledge, only a few works have studied the IRS-aided URLLC networks \cite{Ranjha2020,Walid2020}. In \cite{Ranjha2020}, the authors proposed a novel polytope-based method to solve the overall decoding error probability minimization problem for an IRS-aided URLLC-enabled UAV system by jointly optimizing the passive beamforming, UAV location and channel blocklength. In \cite{Walid2020}, the authors studied the resource allocation problem for IRS-aided multiple-input single-output (MISO) orthogonal frequency division multiple access (OFDMA) multicell networks. However, these two papers focused on the transmission design, the analytical performance of IRS-aided systems is not yet available in the existing literature.

Against the above background, we present a comprehensive performance analysis of the end-to-end IRS-aided URLLC-enabled system in a factory automation scenario. Specifically, the main contributions are summarized as follows:
\begin{enumerate}
  \item We first construct the system model for IRS-aided URLLC in a factory automation scenario. In specific, there is one central control that attempts to send control signals to a remote device. Due to the dense blockage between the transmitter and receiver, the direct signal power is weak. To enhance the communication quality, an IRS is deployed between the transmitter and the receiver, and an alternative transmission link is created to aid the communications. Due to the stringent latency requirement and the small packet size of control signals, the short packet transmission theory should be adopted, which is more complex than conventional Shannon capacity expression.
  \item We have derived the closed-form expressions for the average data rate (ADR) and average decoding error probability (ADEP) under finite channel blocklength for seven cases: 1) Rayleigh fading channel; 2) With direct channel link; 3)  Nakagami-m fading channel; 4) Imperfect phase alignment; 5) Multiple-IRS case; 6) Rician fading channel; 7) Correlated channels. In specific, the distribution of SNR is first approximated as a Gamma distribution by using the moment matching technique, based on which the ADR is derived in closed form. By using the linearization technique, we also derive the approximate closed-form expression of average decoding error probability (ADEP) under finite channel blocklength.
  \item Extensive numerical results obtained from Monte-Carlo simulations   demonstrate the accuracy of the derived results and we also provide insightful analysis. Specifically, there is a roughly fixed gap between Shannon capacity and ADR under finite channel blocklength, which implies that the conventional Shannon capacity overestimates the system performance.
\end{enumerate}

The rest of this paper is organized as follows. Section \ref{system} introduces the system model with IRS-aided point-to-point system along with the short packet theory;  Section \ref{withCSI} is devoted to performance analysis for the simple Rayleigh fading channel. Section \ref{extensions} provides some extensions to six more general cases. Numerical results are presented in Section \ref{simlresult} to validate the accuracy of the derived results. Finally, the conclusion is drawn in Section \ref{conclu}.

\vspace{-0.3cm}\section{System Model and Problem Formulation}\label{system}
\subsection{System  Model}

\begin{figure}
\centering
\includegraphics[width=3.8in]{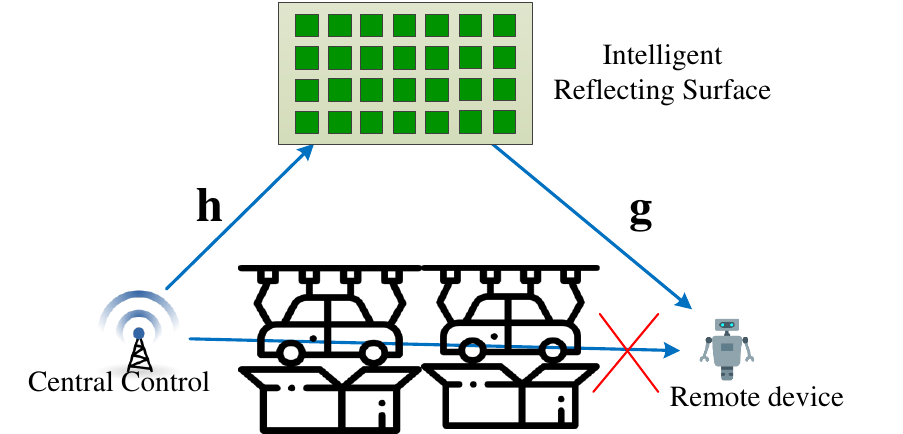}
\caption{Illustration of an IRS-aided automated factory scenario.}
\vspace{-0.3cm}
\label{systemodel}
\end{figure}

As shown in Fig.~\ref{systemodel}, we consider an IRS-aided transmission system for an automated factory scenario, where the central controller transmits command signals with short packet size to one remote device (e.g., a robot or an actuator) \footnote{When there are multiple devices, the orthogonal frequency division multiplexing (OFDM) technique can be adopted, where each device can be allocated with one sub-carrier to ensure the reliability. The following derivations are still applicable.}. Due to the severe blockage caused by huge metallic machines in car manufacturing plants, the channel strength of direct links between the controller and the device is very weak and even lost. The communication link can be established via an IRS installed on the wall or the ceiling of the factory. For the sake of analysis, we assume that the controller and the device are equipped with one single antenna. The IRS is composed of $N$ reflecting phase shifters. The complex channel vectors from the controller to the IRS and from the IRS to the device are denoted by ${\bf{h}}\in \mathbb{C}^{N\times 1}$ and ${\bf{g}}\in \mathbb{C}^{N\times 1}$, respectively. Specifically, the channel vector ${\bf{h}}$ can be represented as ${{\bf{h}}} = {\left[ {h_1, \cdots ,h_N} \right]^{\rm{H}}}$, where $h_n$ denotes the complex channel coefficient from the controller to the $n$-th element of the IRS that follows the distribution of ${\cal{CN}}(0, \alpha)$, where  ${\cal{CN}}(0, \alpha)$ means the  complex Gaussian distribution with zero mean and the variance of $\alpha$. Similarly, we can express ${\bf{g}}$ as ${{\bf{g}}} = {\left[ {g_1, \cdots ,g_N} \right]^{\rm{H}}}$, where $g_n$ follows the distribution of ${\cal{CN}}(0, \beta)$.  In our analysis, both ${\bf{h}}$ and ${\bf{g}}$ are assumed to be available at the transmitter,  which includes the instantaneous realization of ${\bf{h}}$ and ${\bf{g}}$  for the phase shift design at the IRS and the distribution of ${\bf{h}}$ and ${\bf{g}}$  for the performance analysis.

Let us first denote the diagonal reflection-coefficient matrix\footnote{$j$ is the imaginary unit.} at the IRS by
\begin{equation}\label{bggfdt}
  {\bm{\Phi}}  = {\rm{diag}}\left\{ {{e^{j{\phi _1}}}, \cdots ,{e^{j{\phi _n}}}, \cdots ,{e^{j{\phi _N}}}} \right\},
\end{equation}
where ${\phi _n} \in [0,2\pi ]$ is the phase shift of the $n$-th element. By assuming that the controller transmits with fixed power $P$, the received signal at the device is given by
\begin{equation}\label{hggr}
  r = \sqrt P {\bf{g}}^{\rm{H}} {\bm{\Phi}} {{\bf{h}}}x + n,
\end{equation}
where $x\in \mathbb{C}$ is the data information with unit power, and $n\in\mathbb{C}$ is the noise at the user that follows zero mean and variance of ${\sigma ^2}$. Then, the instantaneous SNR at the device is given by
\begin{equation}\label{hdyuju}
  \gamma  = \rho{{{\left| {{\bf{g}}^{\rm{H}}{\bm{\Phi}} {{\bf{h}}}} \right|}^2}},
\end{equation}
where $\rho  = \frac{P}{{{\sigma ^2}}}$.

\vspace{-0.3cm}
\subsection{Short Packet Transmission Theory}

The coding rate for a communication system is defined as the ratio of the number of bits to the total number of channel uses. Shannon capacity is the largest coding rate that can be achieved such that there exists an encoder/decoder pair whose decoding error probability can approach zero when the number of channel uses is sufficiently large \cite{shannon2001mathematical}.
However, for applications in industrial factories, the number of channel uses (channel blocklength) cannot be very large due to the stringent delay requirement. As a result, the decoding error probability will not be equal to zero and has to be reconsidered \cite{hongrentwc,hongrenjsac}.

Let us first denote the maximum decoding error probability as $\varepsilon $. Then, the maximum instantaneous achievable data rate $R$ (bit/s/Hz) \cite{polyanskiy2010channel} of the remote device can be approximated by
 \begin{equation}\label{frsghty}
   R = {\log _2}\left( {1 + \gamma } \right) - \sqrt {\frac{V(\gamma)}{M}} \frac{{{Q^{ - 1}}\left( \varepsilon  \right)}}{{\ln 2}},
 \end{equation}
  where $M$ is the number of channel uses,  $Q^{ - 1}(\cdot)$ is the inverse function of $Q(x) \!=\! \frac{1}{{\sqrt {2\pi } }}\int_x^\infty  {{e^{ - \frac{{{t^2}}}{2}}}dt} $,  and $V(\gamma)$ is the channel dispersion given by $V(\gamma) \!=\! 1 - {(1 + \gamma )^{ - 2}}$. Assume that we require to transmit a packet with $D$ bits within $M$ channel uses. Then, the achievable date rate can be calculated by $R=D/M$, and the corresponding minimum decoding error probability is given by
\begin{equation}\label{dwfaref}
  { \varepsilon }=Q\left( {f\left( {{{\gamma}},{M},D} \right)} \right),
\end{equation}
where $f\!\left( {\gamma ,M,D} \right)\!=\! \ln 2\sqrt {\frac{M}{V(\gamma)}} \left( {{{\log }_2}(1 + \gamma ) \!-\! \frac{D}{M}} \right)$.

\vspace{-0.3cm}
\section{Performance Analysis}\label{withCSI}

Since the CSI is assumed to be available at the BS, the SNR of $\gamma$ in (\ref{hdyuju}) can be rewritten as
\begin{equation}\label{cdarfea}
  \gamma  = \rho\left| {\sum\limits_{n = 1}^N {{g_n}{h_n}{e^{j{\phi _n}}}} } \right|^2.
\end{equation}
With the aid of CSI at the BS, the optimal $\phi_n$ that maximizes the instantaneous $\gamma $ is given by
\begin{equation}\label{derfre}
  {\phi _n^\star} =  - \angle {g_n} - \angle {h_n}.
\end{equation}
Then,  $\gamma $ is given by
\begin{equation}\label{sdwdedwf}
  \gamma  = \rho{\left( {\sum\limits_{n = 1}^N {\left| {{g_n}} \right|\left| {{h_n}} \right|} } \right)^2}.
\end{equation}

Let us define $X \buildrel \Delta \over = {\left( {\sum\nolimits_{n = 1}^N {\left| {{g_n}} \right|\left| {{h_n}} \right|} } \right)^2}$. In the following lemma, we approximate $X$ as a Gamma distribution.


\noindent{\textbf{Lemma 1}}:  Let us denote the first and second moments of the random variable (RV) $X$ by ${u_X} = \mathbb{E}\left\{ X \right\}$ and $u_X^{(2)} = \mathbb{E}\left\{ {{X^2}} \right\}$, respectively. Then, the distribution of $X$ can be approximated as a Gamma distribution with the same first and second moments, and its  shape $\bar k$ and scale $\bar \theta$ are given by
\begin{equation}\label{dewfef}
  \bar k = \frac{{u_X^2}}{{u_X^{(2)} - u_X^2}},\bar \theta  = \frac{{u_X^{(2)} - u_X^2}}{{{u_X}}},
\end{equation}
where $u_X$ and $u_X^{(2)}$ are given in (\ref{grthyth}) and (\ref{jroigjt}), respectively.
The PDF and CDF of   $X$ are respectively approximated  by
\begin{equation}\label{erdewdesg}
  {f_X}(x) = \frac{1}{{\Gamma (\bar k){{\bar \theta }^{\bar k}}}}{x^{\bar k - 1}}{e^{ - \frac{x}{{\bar \theta }}}},{F_X}(x) = \frac{1}{{\Gamma \left( {\bar k} \right)}}\gamma \left( {\bar k,\frac{x}{{\bar \theta }}} \right).
\end{equation}
\emph{\textbf{Proof}}: Please refer to Appendix \ref{lemma1}. \hfill\rule{2.7mm}{2.7mm}

Since $\gamma  = \rho X$, we have $\mathbb{E}\{\gamma\}=\rho \mathbb{E}\{X\}$ and $\mathbb{E}\{\gamma^2\}=\rho^2\mathbb{E}\{X^2\}$. Then, $\gamma$ can  also  be approximated as a Gamma distribution with $k =\bar k$ and $\theta  = \rho \bar \theta $. The PDF and CDF of $\gamma$ are respectively approximated  by
\begin{equation}\label{erdec}
{f_\gamma }(x) = \frac{1}{{\Gamma (k){\theta ^k}}}{x^{k - 1}}{e^{ - \frac{x}{\theta }}},{F_X}(x) = \frac{1}{{\Gamma \left( k \right)}}\gamma \left( {k,\frac{x}{\theta }} \right).
\end{equation}


Based on the above results, we proceed to derive the ADR and  ADEP in the following subsections.
\vspace{-0.3cm}
\subsection{ Average Data Rate}
It is difficult to get the exact expression of the average data rate (ADR). Next, we show the approximate expression of ADR.

 \noindent{\textbf{Lemma 2}}: By using $V(\gamma) = 1-\frac{1}{(\gamma+1)^2} \approx 1$ when $\gamma$ is large, ADR can be approximated as

 \begin{equation}\label{gd8}
 	\begin{aligned}
 		&{\bar R} \approx \frac{\Gamma ({k}-1) \, _2F_2\left(1,1;2,2-{k};\frac{1}{{\theta}}\right)}{{\theta} (\log (2) \Gamma ({k}))}-\frac{{{Q^{ - 1}}\left( \varepsilon  \right)}}{\sqrt{M} \log (2)}+\frac{\psi ^{(0)}({k})+\log ({\theta})}{\log (2)} +\\& \frac{\left(\pi  (-1)^{-{k}} \csc (\pi  {k})\right) \gamma \left({k},-\frac{1}{{\theta}}\right)}{\log (2) \Gamma ({k})},
 	\end{aligned}
 \end{equation}
where $ _pF_q (a;b;z)$ is the generalized hypergeometric function \cite{HypergeometricPFQ}, ${\psi ^{(0)}}(n)$ is the  polygamma function \cite{Poly,gradshteyn2014table}.

\emph{{Proof}}: See Appendix \ref{lemma2}. \hfill\rule{2.7mm}{2.7mm}

%
%

\subsection{Average Decoding Error Probability}

In this subsection, we aim to derive the ADEP by transmitting a packet with fixed size of $D$ within $M$ channel uses. Then, the ADEP is defined as
\begin{equation}\label{cdscdsr}
  \bar \varepsilon  =  \int_0^\infty  {Q\left( {\ln 2\sqrt {\frac{M}{{V(x)}}} \left( {{{\log }_2}(1 + x) - \frac{D}{M}} \right)} \right){f_\gamma }\left( x \right)dx},
\end{equation}
where $f_\gamma(x)$ is given in (\ref{erdec}).

Again, by using the linearization technique to approximate the Q-function, we have the following lemma.

 \noindent{\textbf{Lemma 3}}: The closed-form expression of $\bar \varepsilon$ can be derived as
 \begin{equation}\label{g17}
	\begin{aligned}
		&\bar \varepsilon   \approx \frac{{\theta}^{-{k}}}{2 \Gamma ({k})} \\& \left((2 {\mu} {x_0}+1) \left({x_0}+\frac{1}{2 {\mu}}\right)^{{k}} E_{-{k}}\left(\frac{{x_0}+\frac{1}{2 {\mu}}}{{\theta}}\right)-(2 {\mu} {x_0}-1) \left({x_0}-\frac{1}{2 {\mu}}\right)^{{k}} E_{-{k}}\left(\frac{{x_0}-\frac{1}{2 {\mu}}}{{\theta}}\right)  \right.\\&\left.+{\theta}^{{k}} \left((1+2 {\mu} {x_0}) \gamma \left({k},\frac{{x_0}+\frac{1}{2 {\mu}}}{{\theta}}\right)+(1-2 {\mu} {x_0}) \gamma \left({k},\frac{{x_0}-\frac{1}{2 {\mu}}}{{\theta}}\right)\right)\right),
	\end{aligned}
\end{equation}
where $E_n (z)=\int _1^{\infty }\frac{ e^{-z t }}{t^n} dt$ gives the exponential integral function \cite{ExpIntegralE}.

\emph{{Proof}}: See Appendix \ref{lemma3}. \hfill\rule{2.7mm}{2.7mm}

\section{Extensions to More General Cases}\label{extensions}

In this section, we consider several possible extensions.
\vspace{-0.3cm}
\subsection{The existence of direct channel link}
In this subsection, we consider the case when there is direct channel link between the BS and the remote device, and its channel coefficient is denoted as $h_0$, following the distribution of ${\cal{CN}}(0, \eta)$. Then, the instantaneous SNR at the device is given by
\begin{equation}\label{directlink}
  \gamma  = \rho{{{\left|h_0+{{\bf{g}}^{\rm{H}}{\bm{\Phi}} {{\bf{h}}}} \right|}^2}}=\rho{{{\left|h_0+ {\sum\limits_{n = 1}^N {{g_n}{h_n}{e^{j{\phi _n}}}} } \right|}^2}},
\end{equation}
where $\rho  = \frac{P}{{{\sigma ^2}}}$.

When   CSI is available at the BS, the optimal $\phi_n$ that maximizes the instantaneous $\gamma $ is given by
\begin{equation}\label{ddwdfre}
  {\phi _n^\star} =  - \angle {g_n} - \angle {h_n}+\angle {h_0} .
\end{equation}
Then,  $\gamma $ is given as
\begin{equation}\label{ssefwdedfef}
  \gamma  = \rho{\left( \left| h_0 \right|+{\sum\limits_{n = 1}^N { \left| {{g_n}} \right|\left| {{h_n}} \right|} } \right)^2}=\rho Y,
\end{equation}
where $Y=\left( \left| h_0 \right|+{\sum\nolimits_{n = 1}^N { \left| {{g_n}} \right|\left| {{h_n}} \right|} } \right)^2$.

In the next lemma, we approximate $Y$ as a Gamma distribution.

\noindent\emph{\textbf{Lemma 4}}: The distribution of $Y$ can be approximated as a Gamma distribution, which is characterized by two parameters $\bar k_{Y}$ and $\bar \theta_{Y}$, i.e.,
 \begin{equation}\label{dewfre}
   Y \sim {\rm{Gamma}}\left( {{\bar k_Y},{\bar \theta _Y}} \right),
 \end{equation}
in which the parameters $\bar k_{Y}$ and $\bar \theta_{Y}$ are given by
\begin{equation}\label{sdcawdW}
 \bar k_Y = \frac{{u_Y^2}}{{u_Y^{(2)} - u_Y^2}},\bar \theta_Y  = \frac{{u_Y^{(2)} - u_Y^2}}{{{u_Y}}},
\end{equation}
where $u_Y$ and $u_Y^{(2)}$ are given in (\ref{gdwedwyth}) and (\ref{hehfoji}), respectively.

\emph{{Proof}}: See Appendix \ref{Lemma4}. \hfill\rule{2.7mm}{2.7mm}

 Since $\gamma  = \rho X$, $\gamma$ can  also  be approximated as a Gamma distribution with $k =\bar k_Y$ and $\theta  = \rho \bar \theta_Y $. The PDF and CDF of $\gamma$ are respectively approximated  by
\begin{equation}\label{erdecsewsg}
{f_\gamma }(x) = \frac{1}{{\Gamma (k){\theta ^k}}}{x^{k - 1}}{e^{ - \frac{x}{\theta }}},{F_X}(x) = \frac{1}{{\Gamma \left( k \right)}}\gamma \left( {k,\frac{x}{\theta }} \right).
\end{equation}

\vspace{-0.3cm}
\subsection{Nakagami-$m$ fading channel}

The channel links from BS to the IRS and from the IRS to the device are assumed to experience the Nakagami-$m$ fading. Specifically, we assume $\left| {{h_n}} \right| \sim {\rm{Nakagami}}({m_1},\alpha ), \forall n$, and $\left| {{g_n}} \right| \sim {\rm{Nakagami}}({m_2},\beta ), \forall n$, where $m_1$ and $m_2$ are the parameters that measure the severity of fading,  $\alpha$ and $\beta$ are the average channel power gains.

By using the phase shift in (\ref{derfre}), the SNR $\gamma $ is given by
\begin{equation}\label{ssdwedwedwwf}
  \gamma  = \rho{\left( {\sum\limits_{n = 1}^N {\left| {{g_n}} \right|\left| {{h_n}} \right|} } \right)^2}=\rho Z,
\end{equation}
where $Z=\left( {\sum\nolimits_{n = 1}^N {\left| {{g_n}} \right|\left| {{h_n}} \right|} } \right)^2$.

Define ${\xi _n} = \left| {{g_n}} \right|\left| {{h_n}} \right|$. In the following lemma, we approximate $Z$ as a Gamma distribution.

\noindent\emph{\textbf{Lemma 5}}: The distribution of $Z$ can be approximated as a Gamma distribution, which is characterized by two parameters $\bar k_{Z}$ and $\bar \theta_{Z}$, i.e.,
 \begin{equation}\label{dewfre}
   Z \sim {\rm{Gamma}}\left( {{\bar k_Z},{\bar \theta _Z}} \right),
 \end{equation}
in which the parameters $\bar k_{Z}$ and $\bar \theta_{Z}$ are given by
\begin{equation}\label{sdcawdW}
 \bar k_Z = \frac{{u_Z^2}}{{u_Z^{(2)} - u_Z^2}},\bar \theta_Z  = \frac{{u_Z^{(2)} - u_Z^2}}{{{u_Z}}},
\end{equation}
where $u_Z$ and $u_Z^{(2)}$ are the same as in (\ref{grthyth}) and (\ref{jroigjt}) except that the moments of ${\xi _n}$ are given by
\[\mathbb{E}\{ \xi _n^k\}  = \frac{{\Gamma \left( {{m_1} + {k \mathord{\left/
 {\vphantom {k 2}} \right.
 \kern-\nulldelimiterspace} 2}} \right)\Gamma \left( {{m_2} + {k \mathord{\left/
 {\vphantom {k 2}} \right.
 \kern-\nulldelimiterspace} 2}} \right)}}{{\Gamma \left( {{m_1}} \right)\Gamma \left( {{m_2}} \right)}}{\left( {\frac{{{m_1}{m_2}}}{{\alpha \beta }}} \right)^{ - {k \mathord{\left/
 {\vphantom {k 2}} \right.
 \kern-\nulldelimiterspace} 2}}}.\]

\emph{{Proof}}: The proof is similar to Appendix \ref{lemma1}, which is omitted for simplicity. \hfill\rule{2.7mm}{2.7mm}

Based on Lemma 5, the SNR $\gamma=\rho Z$ is approximated as the Gamma distribution $\gamma \sim {\rm{Gamma}}\left( {{k}, \theta} \right)$, where $k=k_{Z}$ and $\theta=\rho {\theta_{Z}}$. The PDF and CDF of $\gamma$ are given by
\begin{equation}\label{erdewwetsg}
  {f_{{\gamma}}}(x) \approx \frac{1}{{\Gamma (k){\theta ^k}}}{x^{k - 1}}{e^{ - \frac{x}{\theta }}},{F_{{\gamma}}}(x) \approx \frac{1}{{\Gamma \left( k \right)}}\gamma \left( {k,\frac{x}{\theta }} \right).
\end{equation}
\vspace{-0.3cm}
\subsection{Imperfect Phase Alignment}

In this subsection, we consider the case when the phase shift cannot be perfectly aligned with the channel phases due to the channel estimation error or limited quantization bits of the phase shifts of the IRS. In this case, the SNR $\gamma $ is given by \begin{equation}\label{ssefdwdwef}
  \gamma  = \rho {\left| {\sum\limits_{n = 1}^N {\left| {{g_n}} \right|\left| {{h_n}} \right|{e^{j{\omega _n}}}} } \right|^2},
\end{equation}
where $\omega_n$ is the phase alignment error that follows the uniform distribution, i.e., ${\omega _n} \sim U\left( { - \Delta ,\Delta } \right)$, and $\Delta$ is a constant characterizing the phase error level. When $\Delta=0$, it reduces to the perfect CSI case in Section \ref{withCSI}.

Define  $G\buildrel \Delta \over = {\left| {\sum\limits_{n = 1}^N {\left| {{g_n}} \right|\left| {{h_n}} \right|{e^{j{\omega _n}}}} } \right|^2} $.  In the following lemma, we approximate $G$  as a Gamma distribution.

\noindent\emph{\textbf{Lemma 6}}: The distribution of $G$ can be approximated as a Gamma distribution, which is characterized by two parameters $\bar k_{G}$ and $\bar \theta_{G}$, i.e.,
 \begin{equation}\label{ddfedwdewre}
   G \sim {\rm{Gamma}}\left( {{\bar k_{G}},{\bar \theta_{G}}} \right),
 \end{equation}
in which the parameters $\bar k_{G}$ and $\bar \theta_{G}$ are given by
\begin{equation}\label{sdfedewderdW}
  \bar k_{G}= \frac{{u_{G}^2}}{{u_{G}^{(2)} - u_{G}^2}},\bar \theta_{G} = \frac{{u_{G}^{(2)} - u_{G}^2}}{{{u_{G}}}},
\end{equation}
where $u_{G}$ and $u_{G}^{(2)}$ are given in (\ref{grtdwdwedyth}) and (\ref{dewdwegjt}), respectively.

\emph{{Proof}}: See Appendix \ref{Lemma6}. \hfill\rule{2.7mm}{2.7mm}

Based on Lemma 6, the SNR $\gamma=\rho G$ is approximated as the Gamma distribution $\gamma \sim {\rm{Gamma}}\left( {{k}, \theta} \right)$, where $k=\bar k_{G}$ and $\theta=\rho {\bar \theta_{G}}$. The PDF and CDF of $\gamma$ are given by
\begin{equation}\label{erdewwetsg}
  {f_{{\gamma}}}(x) \approx \frac{1}{{\Gamma (k){\theta ^k}}}{x^{k - 1}}{e^{ - \frac{x}{\theta }}},{F_{{\gamma}}}(x) \approx \frac{1}{{\Gamma \left( k \right)}}\gamma \left( {k,\frac{x}{\theta }} \right).
\end{equation}

\vspace{-0.3cm}
\subsection{Multiple-IRS Case}

In this subsection, we consider the case when there are multiple IRSs, the number of which is $I$. The IRS are distributed far away such that the channels to/from IRSs are uncorrelated, which can increase the diversity. Denote the channel from the controller to the $i$-th IRS and that from the $i$-th IRS to the user as ${\bf{h}}_i$ and ${\bf{g}}_i$, respectively. Each element of ${\bf{h}}_i$ and ${\bf{g}}_i$ follows the distribution of ${\cal{CN}}(0, \alpha_i)$ and ${\cal{CN}}(0, \beta_i)$, respectively. Then, the SNR of $\gamma$  can be rewritten as
\begin{equation}\label{cddwdewdwea}
  \gamma  = \rho {\left| {\sum\limits_{i = 1}^I {\sum\limits_{n = 1}^N {{g_{i,n}}{h_{i,n}}{e^{j{\phi _{i,n}}}}} } } \right|^2},
\end{equation}
where $h_{i,n}$ and $g_{i,n}$ are the $n$-th element of ${\bf{h}}_i$ and ${\bf{g}}_i$.

With the aid of CSI at the controller, the optimal $\phi_{i,n}$ that maximizes the instantaneous $\gamma $ is given by
\begin{equation}\label{dewedwede}
  {\phi _{i,n}^\star} =  - \angle {g_{i,n}} - \angle {h_{i,n}}.
\end{equation}
Then,  $\gamma $ is given as
\begin{equation}\label{sdwwdewdewf}
  \gamma  = \rho {\left( {\sum\limits_{i = 1}^I {\sum\limits_{n = 1}^N {\left| {{g_{i,n}}} \right|\left| {{h_{i,n}}} \right|} } } \right)^2}.
\end{equation}

Define ${\xi _{i,n}} = \left| {{g_{i,n}}} \right|\left| {{h_{i,n}}} \right|$ and $\xi_{i}  = \sum\nolimits_{n = 1}^N {{\xi _{i,n}}}$, then the SNR can be rewritten as $\gamma  = \rho {\left( {\sum\nolimits_{i = 1}^I {{\xi _i}} } \right)^2}=\rho U$, where $U=\left( {\sum\nolimits_{i = 1}^I {{\xi _i}} } \right)^2$.

In the following lemma, we approximate $U$  as a Gamma distribution.

\noindent\emph{\textbf{Lemma 7}}: The distribution of $U$ can be approximated as a Gamma distribution, which is characterized by two parameters $\bar k_{U}$ and $\bar \theta_{U}$, i.e.,
 \begin{equation}\label{ddfedwfwswddewre}
   U \sim {\rm{Gamma}}\left( {{\bar k_{U}},{\bar \theta_{U}}} \right),
 \end{equation}
in which the parameters $k_{U}$ and $\theta_{U}$ are given by
\begin{equation}\label{sdfedeedwdewerdW}
  \bar k_{U}= \frac{{u_{U}^2}}{{u_{U}^{(2)} - u_{U}^2}},\bar \theta_{U} = \frac{{u_{U}^{(2)} - u_{U}^2}}{{{u_{U}}}},
\end{equation}
where $u_{U}$ and $u_{U}^{(2)}$ are given in (\ref{fjrtogt}) and (\ref{vjhodn}), respectively.

\emph{{Proof}}: See Appendix \ref{Lemma7}. \hfill\rule{2.7mm}{2.7mm}

Based on Lemma 7, the SNR $\gamma=\rho U$ is approximated as the Gamma distribution $\gamma \sim {\rm{Gamma}}\left( {{k}, \theta} \right)$, where $k=\bar k_{U}$ and $\theta=\rho {\bar \theta_{U}}$. The PDF and CDF of $\gamma$ are given by
\begin{equation}\label{erdTGETGDEetsg}
  {f_{{\gamma}}}(x) \approx \frac{1}{{\Gamma (k){\theta ^k}}}{x^{k - 1}}{e^{ - \frac{x}{\theta }}},{F_{{\gamma}}}(x) \approx \frac{1}{{\Gamma \left( k \right)}}\gamma \left( {k,\frac{x}{\theta }} \right).
\end{equation}
\vspace{-0.3cm}
\subsection{Rician Fading Channel}
In this subsection, we consider the Rician channel model.  Specifically, we assume $\left| {{h_n}} \right| \sim {\rm{Rician}}(\alpha_1,\beta_1), \forall n$, the PDF of which is given by
\begin{equation}\label{ewewkkk}
  {f_{\left| {{h_n}} \right|}}\left( x \right) = \frac{x}{{{\alpha_1^2}}}\exp \left( { - \frac{{{x^2} + {\beta_1^2}}}{{2{\alpha_1 ^2}}}} \right){I_0}\left( {\frac{{x\beta_1 }}{{{\alpha_1^2}}}} \right),
\end{equation}
where $I_0(z)$ is the modified Bessel function of the first kind with order zero. In the above Rician fading, the shape parameter $K = \frac{{\beta _1^2}}{{2\alpha_1^2}}$ denotes  the ratio of the power contributions by line-of-sight path to the remaining multipaths, and  the Scale parameter  $\Omega  = 2\alpha _1^2 + \beta _1^2$ is the total power received in all paths. In addition, we assume $\left| {{g_n}} \right| \sim {\rm{Rician}}(\alpha_2,\beta_2), \forall n$.

By using the phase shift in (\ref{derfre}), the SNR $\gamma $ is given by
\begin{equation}\label{sswdewdfewf}
  \gamma  = \rho{\left( {\sum\limits_{n = 1}^N {\left| {{g_n}} \right|\left| {{h_n}} \right|} } \right)^2}=\rho \left( {\sum\limits_{n = 1}^N {{\xi _n}} }\right)^2=\rho V,
\end{equation}
where we have defined ${\xi _n} = \left| {{g_n}} \right|\left| {{h_n}} \right|$ and $V=\left( {\sum\nolimits_{n = 1}^N {{\xi _n}} }\right)^2$.
Define ${\xi _n} = \left| {{g_n}} \right|\left| {{h_n}} \right|$ and $\xi  = \sum\nolimits_{n = 1}^N {{\xi _n}}$. In the following lemma, we approximate $V$ as a Gamma distribution.

\noindent\emph{\textbf{Lemma 8}}: The distribution of $V$ can be approximated as a Gamma distribution, which is characterized by two parameters $\bar k_{V}$ and $\bar \theta_{V}$, i.e.,
 \begin{equation}\label{ddfdedewdfre}
   V \sim {\rm{Gamma}}\left( {{\bar k_{V}},{\bar \theta_{V}}} \right),
 \end{equation}
in which the parameters $\bar k_{V}$ and $\bar \theta_{V}$ are given by
\begin{equation}\label{sdfdewerdW}
 \bar k_{V}= \frac{{u_{V}^2}}{{u_{V}^{(2)} - u_{V}^2}},\bar \theta_{V} = \frac{{u_{V}^{(2)} - u_{V}^2}}{{{u_{V}}}},
\end{equation}
where $u_{V}$ and $u_{V}^{(2)}$ are the same as in (\ref{grthyth}) and (\ref{jroigjt}) except that the moments of ${\xi _n}$ are given by \cite{Ricianfunchde}
\[ u_{{\xi _n}}^{(k)} = \alpha _1^k{2^{k/2}}\Gamma \left( {1 + {k \mathord{\left/
 {\vphantom {k 2}} \right.
 \kern-\nulldelimiterspace} 2}} \right){L_{k/2}}\left( { - {{\beta _1^2} \mathord{\left/
 {\vphantom {{\beta _1^2} {2\alpha _1^2}}} \right.
 \kern-\nulldelimiterspace} {2\alpha _1^2}}} \right),\]
 where  ${L_q}(x)$ denotes a Laguerre polynomial ${L_{q}}(z) = {}_1{F_1}\left( { - q;1;z} \right)$.

\emph{{Proof}}: The proof is similar to Appendix \ref{lemma1}, which is omitted for simplicity. \hfill\rule{2.7mm}{2.7mm}

Then, the random variable $\xi$ follows the Gamma distribution with parameters equal to $Nk_{\xi _n}$ and $\theta_{\xi _n}$. The following steps are the same as those in Section \ref{withCSI}.

\subsection{Correlated Channels}
In this subsection, we consider the correlated channels at the IRS. Specifically, the channels follow the following distribution:
\begin{equation}\label{juyjuyt}
  {\bf{h}}\sim \cal{CN}(\bf{0}, \alpha \bf{R_{\rm{ci}}}), {\bf{g}} \sim \cal{CN}(\bf{0}, \beta \bf{R_{\rm{id}}}),
\end{equation}
where $\bf{R_{\rm{ci}}}$ and $\bf{R_{\rm{id}}}$ are the covariance matrices.

Assume that IRS with $N=N_{\rm{H}}N_{\rm{V}}$ reflecting elements is a two-dimensional rectangular surface,  which  has $N_{\rm{H}}$ elements per row and $N_{\rm{V}}$ elements per column. Each element is assumed to have size of ${d_{\rm{H}}} \times {d_{\rm{V}}}$, where $d_{\rm{H}}$ and $d_{\rm{V}}$ are the horizontal width and vertical height, respectively. We consider the local spherical coordinate system as in \cite{Emilwcl2021}, and the location of the $n$th element with respect to the origin is ${{\bf{u}}_n} = {\left[ {0,i(n){d_{\rm{H}}},j(n){d_{\rm{V}}}} \right]^{\rm{T}}}$, where $i(n)={\rm{mod}}(n-1,N_{\rm{H}})$ and $j(n)= \left\lfloor {{{\left( {n - 1} \right)} \mathord{\left/
 {\vphantom {{\left( {n - 1} \right)} {{N_{\rm{H}}}}}} \right.
 \kern-\nulldelimiterspace} {{N_{\rm{H}}}}}} \right\rfloor $. Here, ${\rm{mod}}(\cdot,\cdot)$ denotes the modulus operation and $\left\lfloor \cdot \right\rfloor$ is the truncation operation. Then, we have $\bf{R_{\rm{ci}}}=\bf{R_{\rm{id}}}=\bf{R}$, and the element of which is given by
 \begin{equation}\label{hjtohjiy}
   {\left[ {\bf{R}} \right]_{n,m}} = {\rm{sinc}}\left( {\frac{{2\left\| {{{\bf{u}}_n} - {{\bf{u}}_m}} \right\|}}{\lambda }} \right),n,m = 1, \cdots ,N
 \end{equation}
where ${\rm{sinc}}(x) = {{\sin \left( {\pi x} \right)} \mathord{\left/
 {\vphantom {{\sin \left( {\pi x} \right)} {\left( {\pi x} \right)}}} \right.
 \kern-\nulldelimiterspace} {\left( {\pi x} \right)}}$ is the sinc function.

Based on the above system model, we then analyze the PDF of the SNR $\gamma  = \rho{{{\left| {{\bf{g}}^{\rm{H}}{\bm{\Phi}} {{\bf{h}}}} \right|}^2}}$ when only the statistic covariance matrix is available. Here, we fix the value of the phase shift matrix of the IRS at ${\bm{\Phi}}$. In the following lemma, we approximate $\gamma$ as a Gamma distribution.

\noindent\emph{\textbf{Lemma 9}}: The distribution of $\gamma$ can be approximated as a Gamma distribution, which is characterized by two parameters $k_{\gamma}$ and $\theta_{\gamma}$, i.e.,
 \begin{equation}\label{ddfddwewdwdfre}
   \gamma \sim {\rm{Gamma}}\left( {{k_{\gamma}},{\theta_{\gamma}}} \right),
 \end{equation}
in which the parameters $k_{\gamma}$ and $\theta_{\gamma}$ are given by
\begin{equation}\label{sdfdedwewerdW}
  k_{\gamma}= \frac{{u_{\gamma}^2}}{{u_{\gamma}^{(2)} - u_{\gamma}^2}},\theta_{\gamma} = \frac{{u_{\gamma}^{(2)} - u_{\gamma}^2}}{{{u_{\gamma}}}},
\end{equation}
where $u_{\gamma}$ and $u_{\gamma}^{(2)}$ are given in (\ref{HJYOTIJIO}) and (\ref{fehfir}), respectively.

\emph{{Proof}}: See Appendix \ref{Lemma9}. \hfill\rule{2.7mm}{2.7mm}

\vspace{-0.3cm}
\section{Simulation Results}\label{simlresult}

In this section, numerical results are provided to verify the accuracy of our derived results. For illustration purposes, the simulation parameters are set as follows: $\alpha=\beta= 1$ and $M=200$. For the case of ADR, $\varepsilon$ is set to $\varepsilon=10^{-6}$, while for the case of ADEP, the number of bits is set to $D=100$. The other parameters are specified in each simulation figure. The curve labelled as `Simulation' is obtained by averaging over 10000 randomly and uniformly generating channels.

\vspace{-0.3cm}
\subsection{Rayleigh Fading Channel}

 \begin{figure}[htbp]
	\centering
	\subfigure[$N=50$]{
		\includegraphics[width=0.46\textwidth]{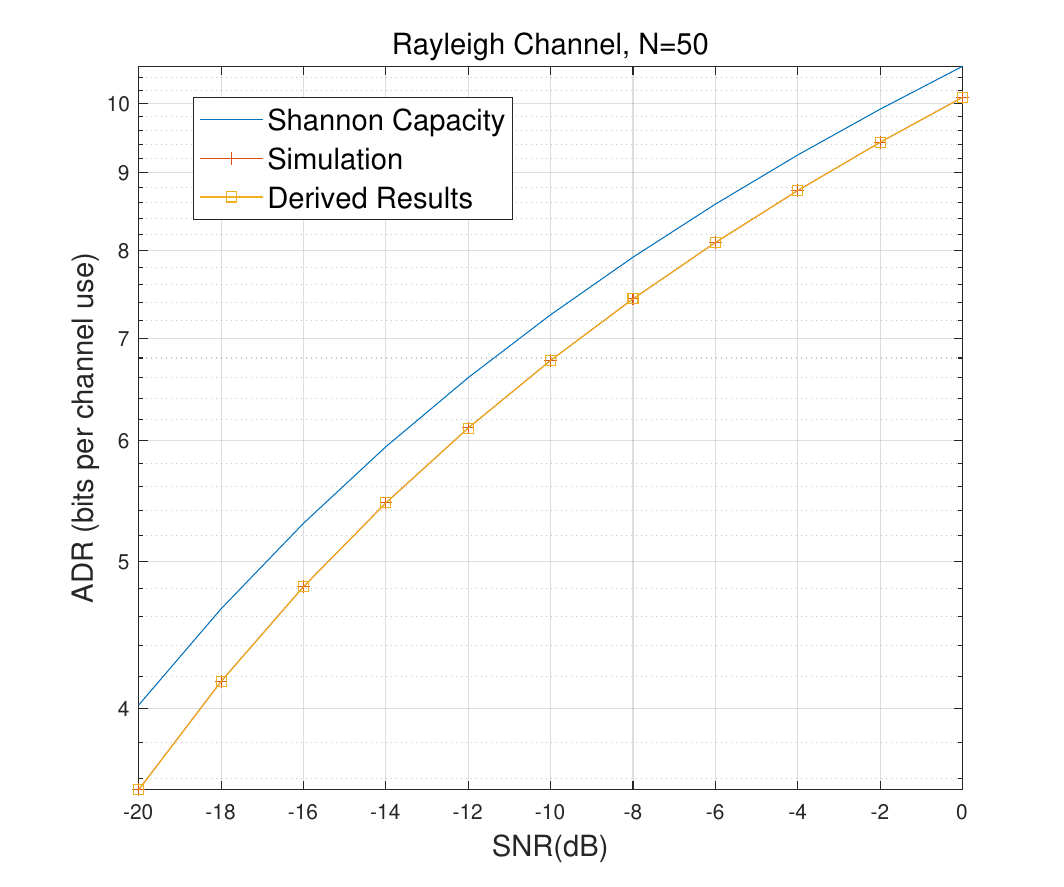}
	}
	\subfigure[$N=100$]{
		\includegraphics[width=0.46\textwidth]{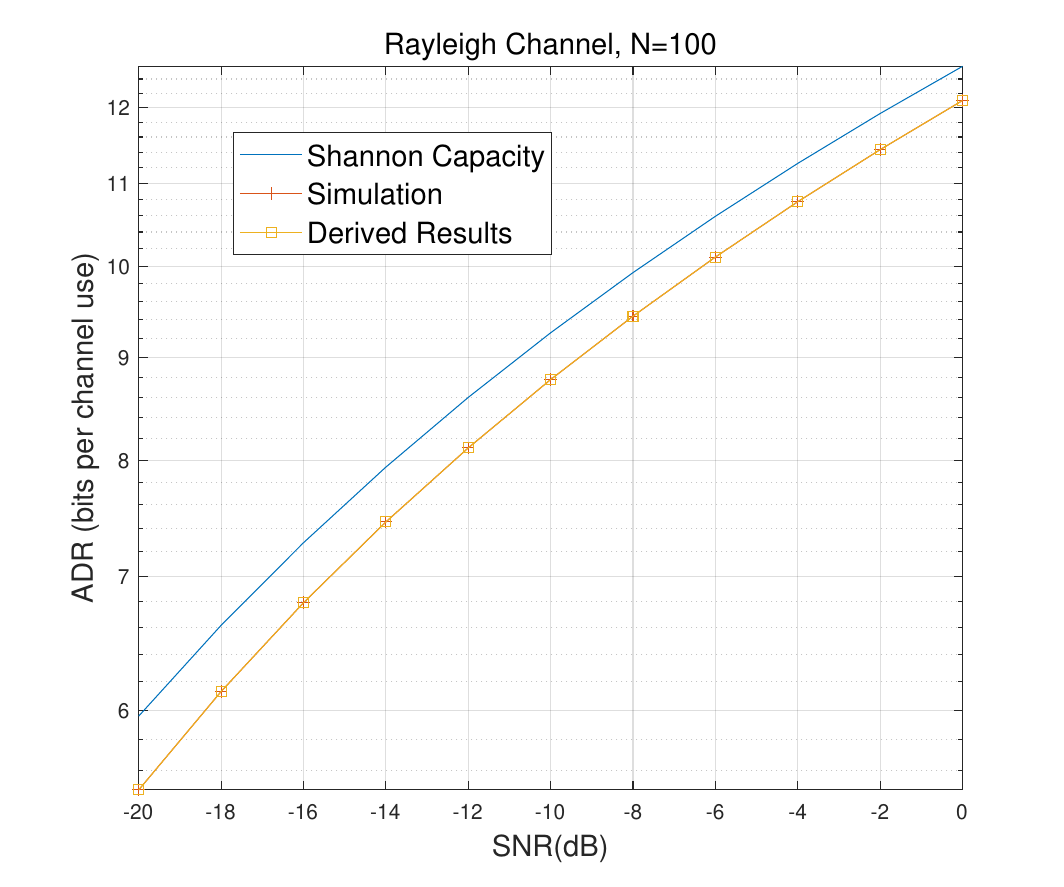}
	}
	\caption{
		ADR versus SNR $N=50$ and $N=100$  for the case of Rayleigh channel.
	} \label{fig1}
\end{figure}

 \begin{figure}[htbp]
	\centering
	\subfigure[$N=50$]{
		\includegraphics[width=0.46\textwidth]{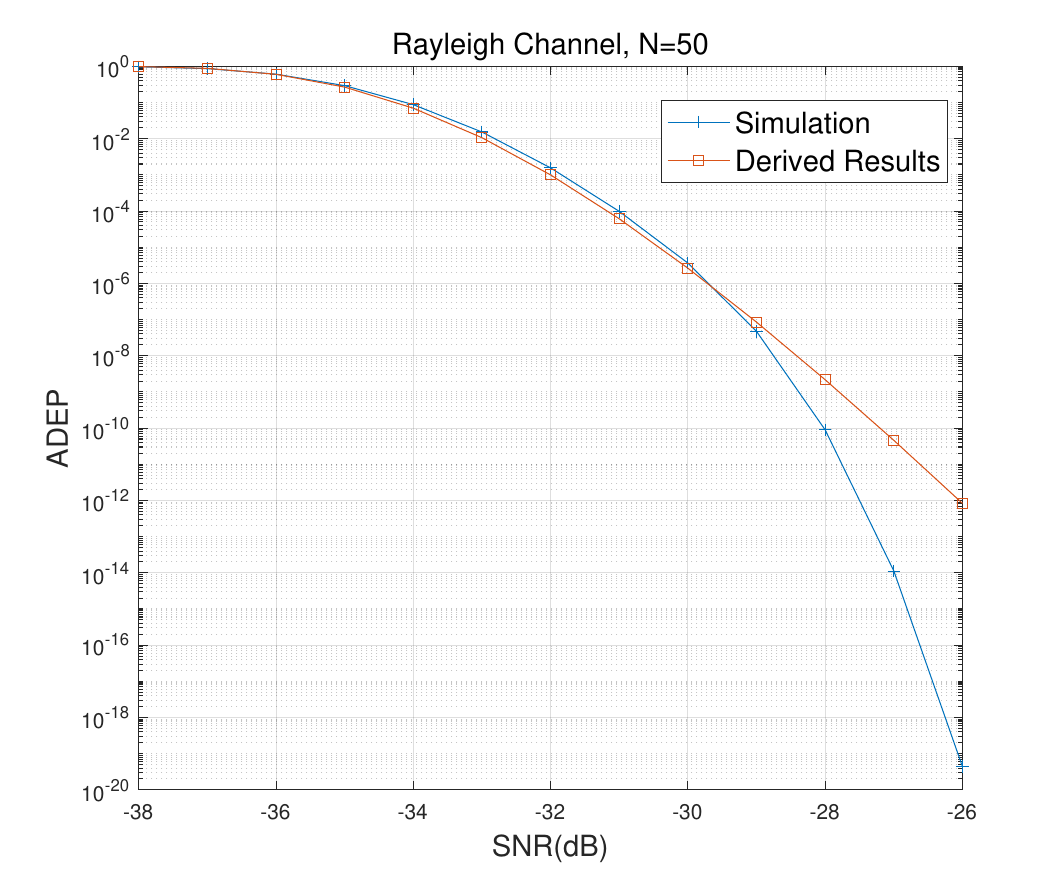}
	}
	\subfigure[$N=100$]{
		\includegraphics[width=0.46\textwidth]{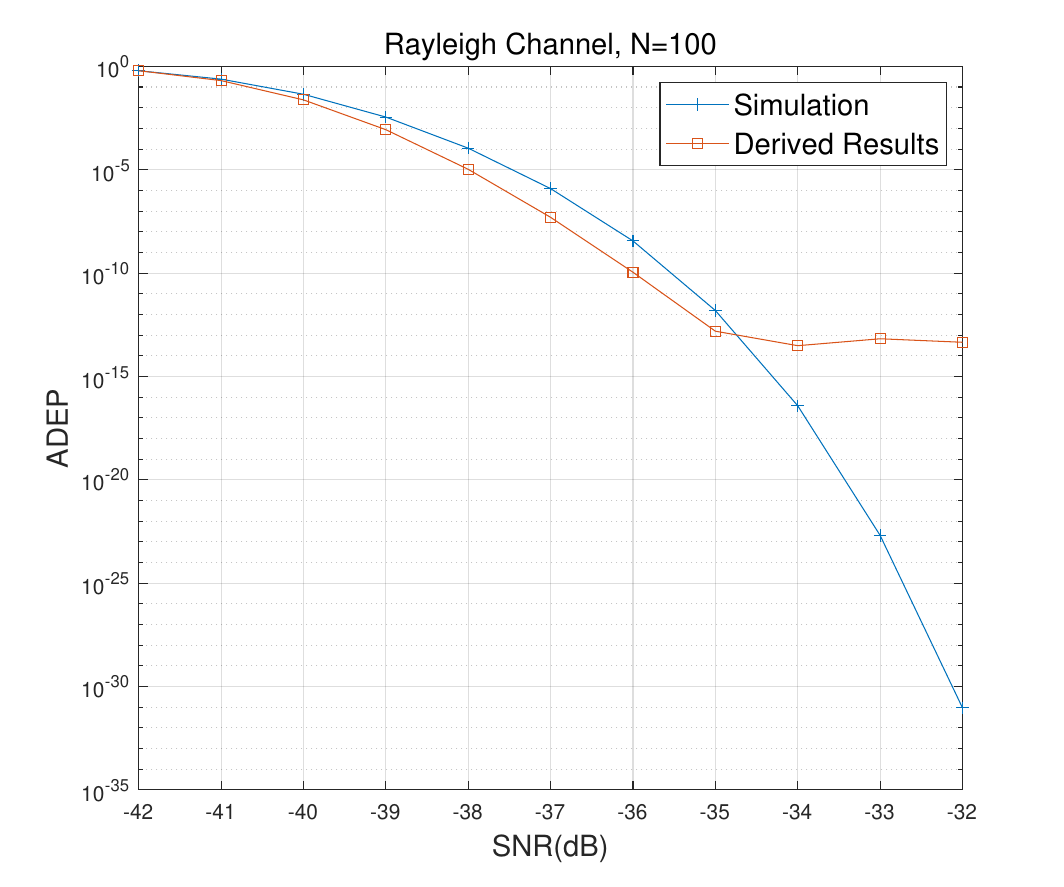}
	}
	\caption{
		ADEP versus SNR $N=50$ and $N=100$  for the case of Rayleigh channel.
	} \label{fig2}
\end{figure}

In this subsection, we consider the Rayleigh fading channel.

In Fig. \ref{fig1}, the ADR versus the SNR (in dB) for two different values of $N$. Here, the SNR in the figures is defined as $10{\log _{10}}\rho$, where $\rho  = \frac{P}{{{\sigma ^2}}}$. The curves labelled as `Derived Results' are obtained  in (\ref{gd8}). For comparison purposes, we also provide the curse based on Shannon's Capacity that can serve as the performance upper bound. As expected, the ADR increases with SNR, and larger $N$ yields high ADR due to higher passive beamforming gain provided by the IRS. It is also observed from Fig. \ref{fig1} that the derived results have a perfect agreement with the simulation results for various SNR values and different values of $N$. In addition, there is a constant gap of roughly 0.5 bits per channel use between the Shannon's capacity and the derived results under the short packet capacity theory.

Fig. \ref{fig2} depicts the ADEP versus the SNR under two different values of $N$. The curve corresponding to `Derived Results' is obtained by using (\ref{g17}). One can observe from Fig. \ref{fig2} that when SNR is smaller than -30 dB for the case of $N=50$ and smaller than -35 dB for the case of $N=100$, the derived results coincide with the simulation results, which confirms the accuracy of the derived results and the  linearization  technique. However, when continuing to increase the SNR value, there is some gap between the derived results and simulation results, and this gap increases with the SNR values. The main reason for this phenomenon is that the value of  Q-function will be approximated as zero when the SNR is very large as shown in
the third approximation result in (\ref{fregtr}). However, the derived results are very accurate when the ADEP is larger than $10^{-6}$, which generally falls the reliability requirements for most of the URLLC applications.
\vspace{-0.3cm}
\subsection{With Direct Channel Link}

\begin{figure}[htbp]
	\centering
	\subfigure[$N=50$]{
		\includegraphics[width=0.46\textwidth]{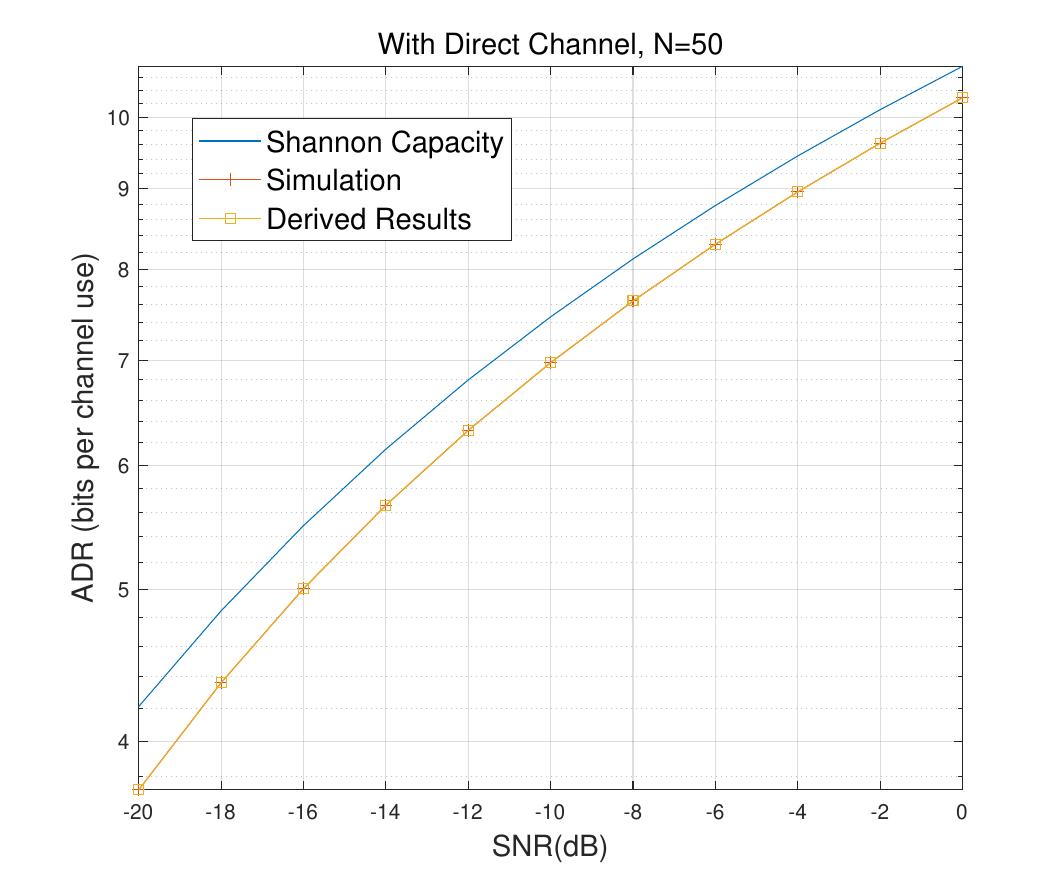}
	}
	\subfigure[$N=100$]{
		\includegraphics[width=0.46\textwidth]{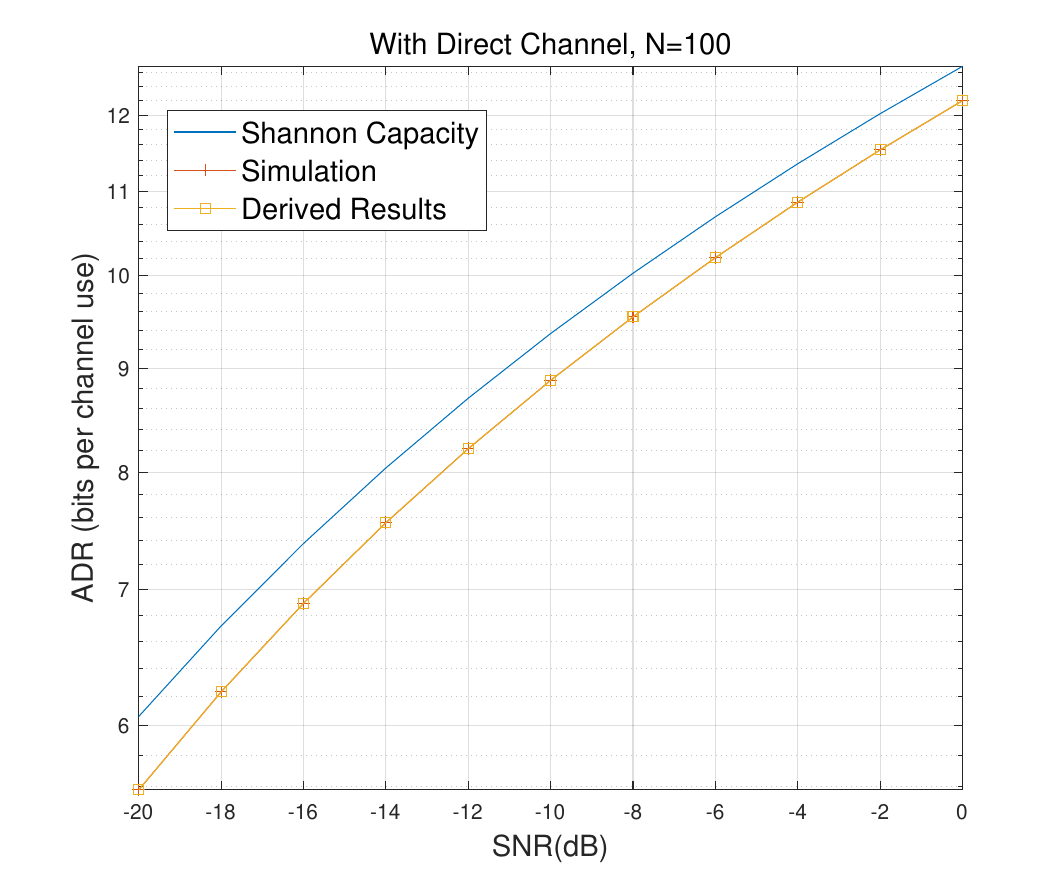}
	}
	\caption{ADR versus SNR  for the case with direct channel.
	} \label{fig3}
\end{figure}

 \begin{figure}[htbp]
	\centering
	\subfigure[$N=50$]{
		\includegraphics[width=0.46\textwidth]{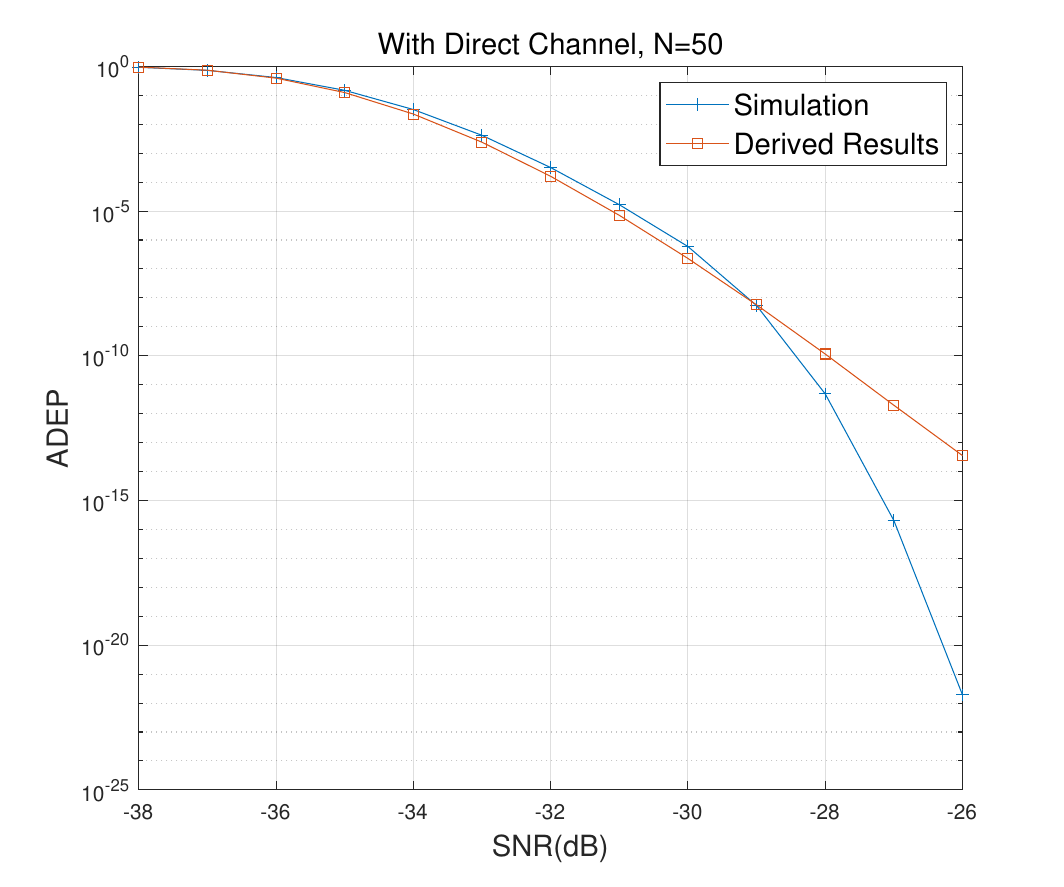}
	}
	\subfigure[$N=100$]{
		\includegraphics[width=0.46\textwidth]{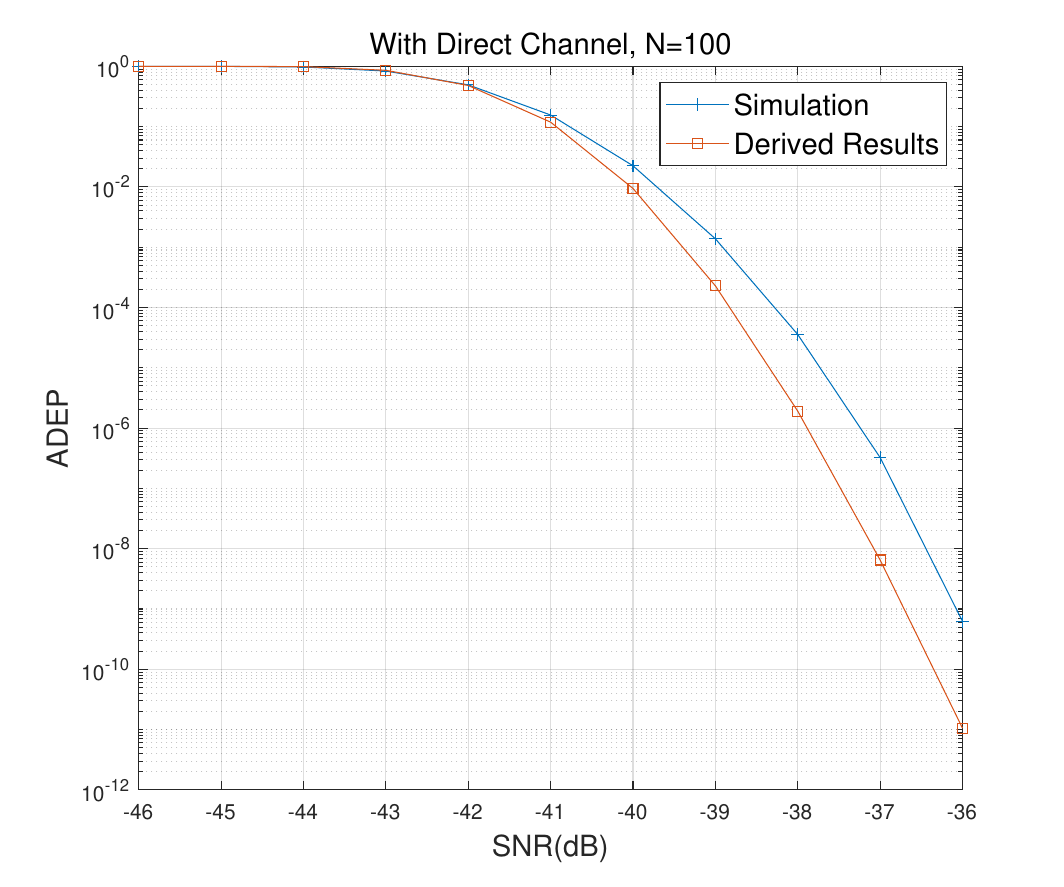}
	}
	\caption{
		ADEP versus SNR  for the case with direct channel.
	} \label{fig4}
\end{figure}
In this subsection, we consider the case when there is a direct channel link between the BS and the device. The channel gain of the direct channel is assumed to be $\eta=10$.

In Fig. \ref{fig3}, the ADR is shown as a function of SNR for two different values of $N$. The similar trend to Fig. \ref{fig1} is observed in Fig. \ref{fig3}. In particular, the ADR increases with the values of SNR, and there is a roughly constant gap of 0.5 bit per channel between the Shannon capacity and that of the ADR based on short packet capacity theory. In addition, a larger value of $N$ produces a higher ADR. \textcolor[rgb]{0.00,0.50,1.00}{By comparing Fig.~\ref{fig3}-(a) with Fig.~\ref{fig1}-(a), the ADR achieved by the case with direct link has a performance gain of 0.2 bit per channel use over the case without direct link when the SNR is equal to -16 dB. However, by  comparing the case of $N=50$ with the case of $N=100$ when SNR is equal to -10 dB, we can find 2 bits per channel use can be achieved, which demonstrate the efficiency of using more reflecting elements.}

Fig.~\ref{fig4} illustrates the ADEP versus SNR for different values of $N$. From Fig.~\ref{fig4}-(a), we can observe that when the SNR is smaller than -30 dB, the derived results are consistent with the simulation results, and can achieve the ADEP as low as $10^{-7}$, which is sufficient for some URLLC applications. However, in the high SNR regime, there is a gap between the derived results and the simulation results, which is mainly due to the approximation error of the linearization technique  in the high SNR regime. For Fig.~\ref{fig4}-(b),  the similar decreasing trend between the `Derived results' and  simulation results is observed, and thus the derived results can be used for revealing the diversity order of the ADEP.
\vspace{-0.3cm}
\subsection{Nakagami-$m$ Fading Channel}

\begin{figure}[htbp]
	\centering
	\subfigure[$N=50$]{
		\includegraphics[width=0.46\textwidth]{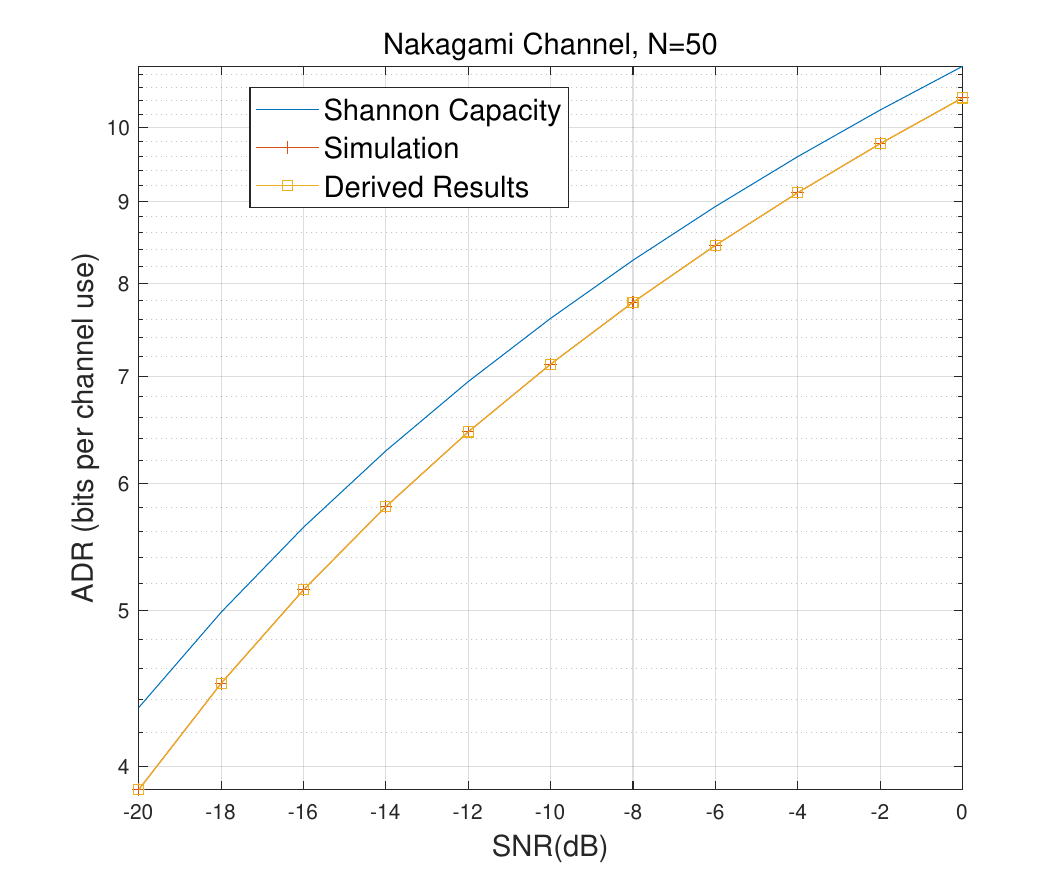}
	}
	\subfigure[$N=100$]{
		\includegraphics[width=0.46\textwidth]{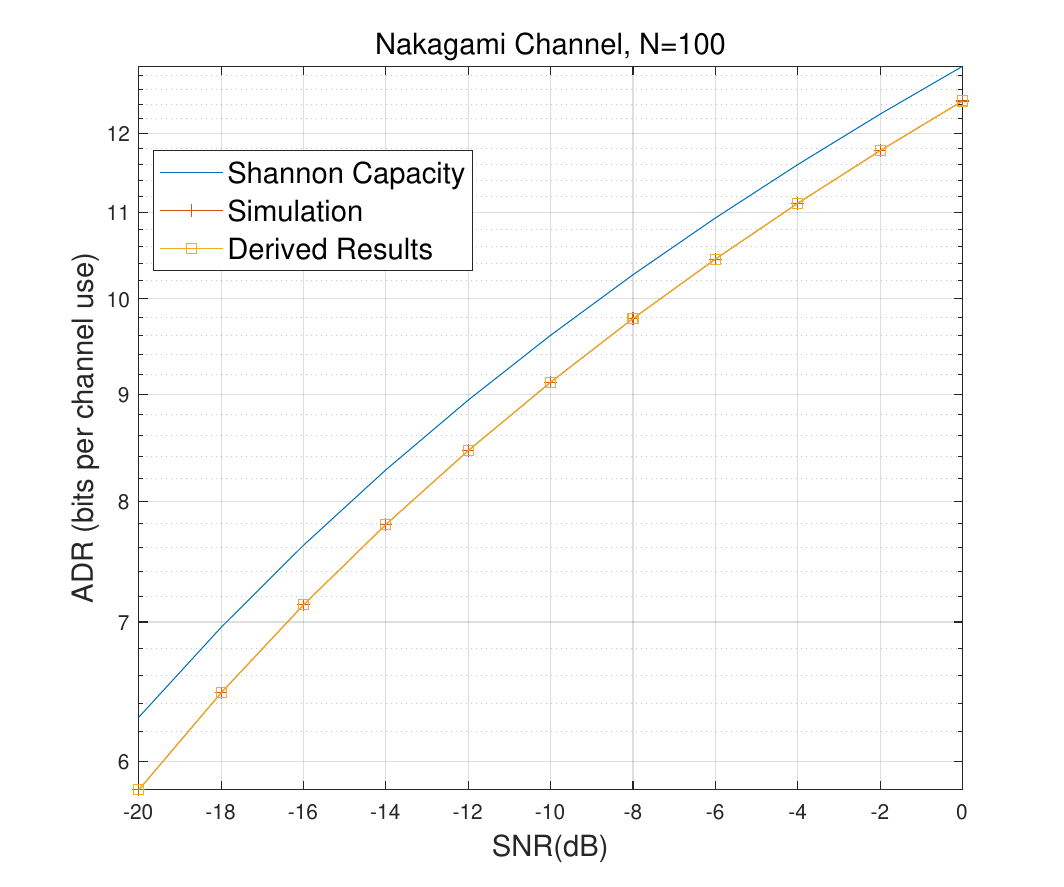}
	}
	\caption{ADR versus SNR  for the case  of Nakagami channel.
	} \label{fig5}
\end{figure}

 \begin{figure}[htbp]
	\centering
	\subfigure[$N=50$]{
		\includegraphics[width=0.44\textwidth]{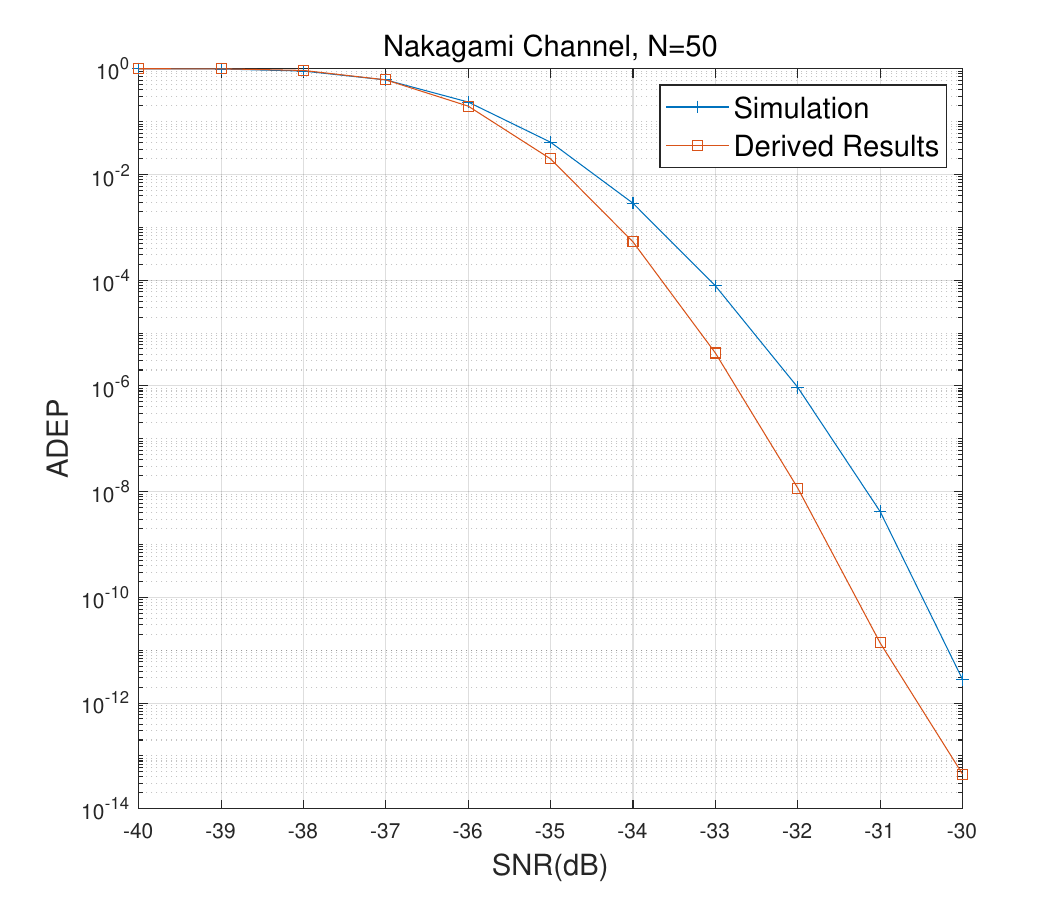}
	}
	\subfigure[$N=80$]{
		\includegraphics[width=0.49\textwidth]{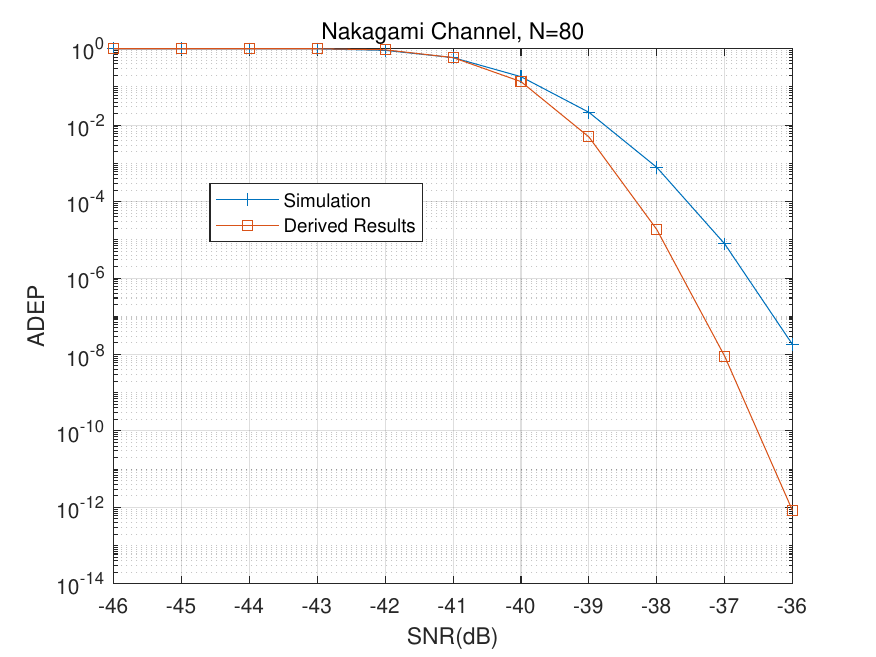}
	}
	\caption{ADEP versus SNR  for the case  of Nakagami channel.
	} \label{fig6}
\end{figure}

In this subsection, we consider the case of Nakagami-$m$ fading channel. The  parameters of $m_1$ and $m_2$ that measure the severity of fading are set as $m_1=2$ and $m_2=2$, respectively.

In Fig.~\ref{fig5}, we illustrate the ADR versus the SNR for two different values of $N$. For comparison, the performance of Shannon Capacity is also shown in the figure. It is observed from Fig.~\ref{fig5} that for both values of $N$, the Shannon capacity achieves high ADR than that based on short packet theory, and this gain decreases with the SNR. This shows that the conventional Shannon capacity will overestimate the system performance, and short packet capacity theory should be adopted to measure the system performance for URLLC applications. \textcolor[rgb]{0.00,0.50,1.00}{By comparing Fig.~\ref{fig5}-(a) with Fig.~\ref{fig1}-(a), we can find that the ADR achieved under the Nakagami-$m$ fading channel is slightly higher than that under the Rayleigh Channel. In addition, the performance gain of 2 bits per channel use can be  achieved when increasing $N$ from 50 to 100, which shows the benefits of using more reflecting elements.}

In Fig.~\ref{fig6}, we illustrate the ADEP versus SNR for different values of $N$. It is again observed that the derived results are accurate when the SNR is very low, and there is  approximation error when the SNR is large due to the approximation error of the linearization technique.

\vspace{-0.3cm}
\subsection{Imperfect Phase Alignment}

\begin{figure}[htbp]
	\centering
	\subfigure[$N=50$]{
		\includegraphics[width=0.46\textwidth]{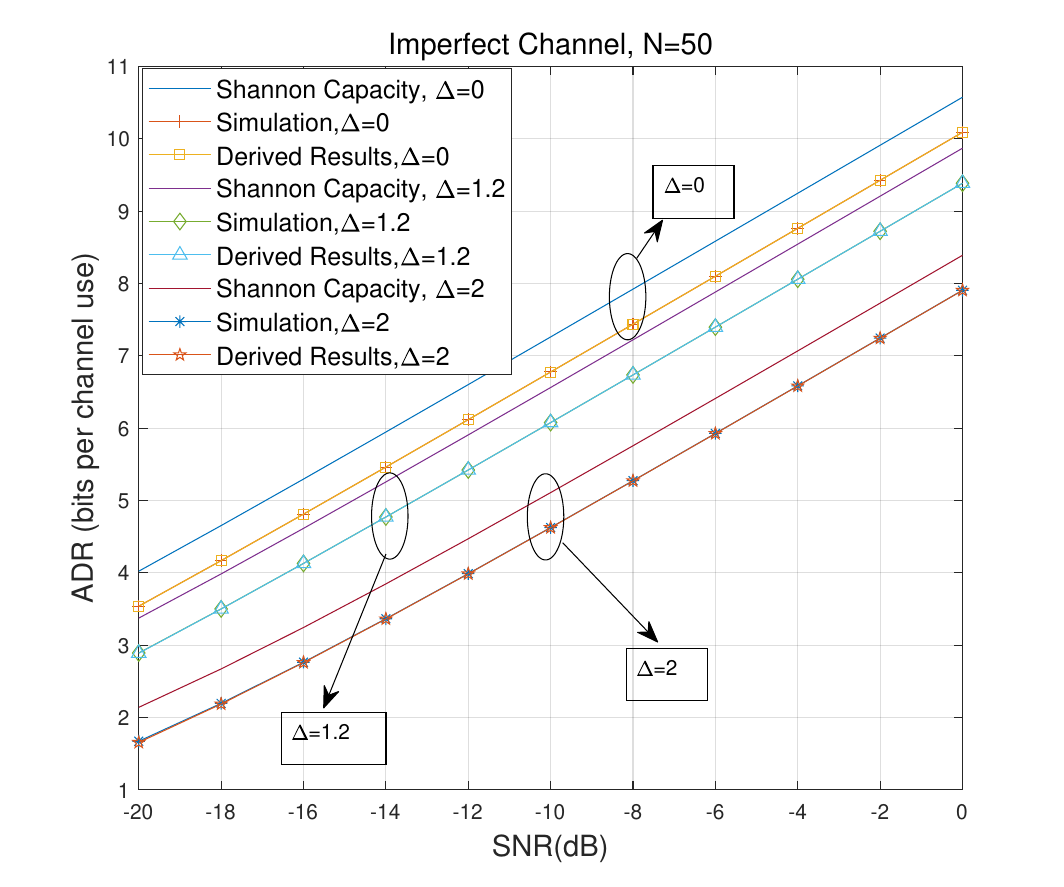}
	}
	\subfigure[$N=100$]{
		\includegraphics[width=0.46\textwidth]{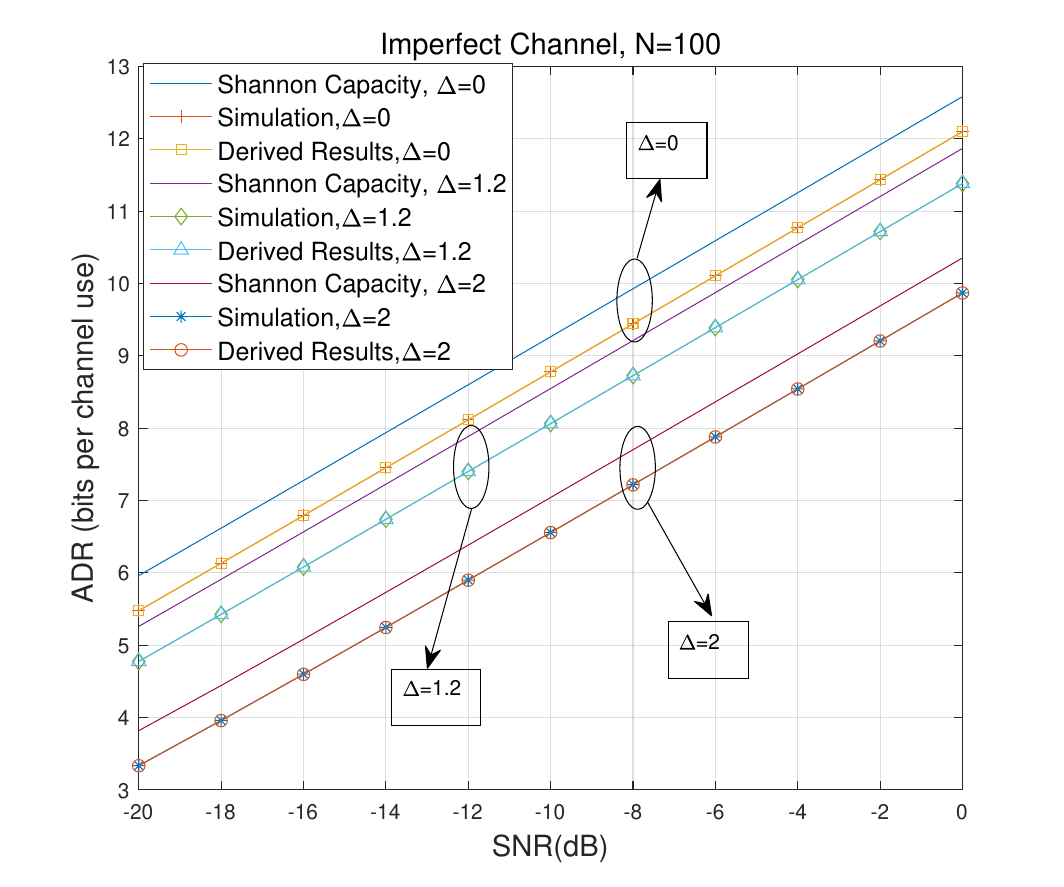}
	}
	\caption{ADR versus SNR  for the case  of imperfect phase alignment.
	} \label{fig7}
\end{figure}

 \begin{figure}[htbp]
	\centering
	\subfigure[$N=50$]{
		\includegraphics[width=0.46\textwidth]{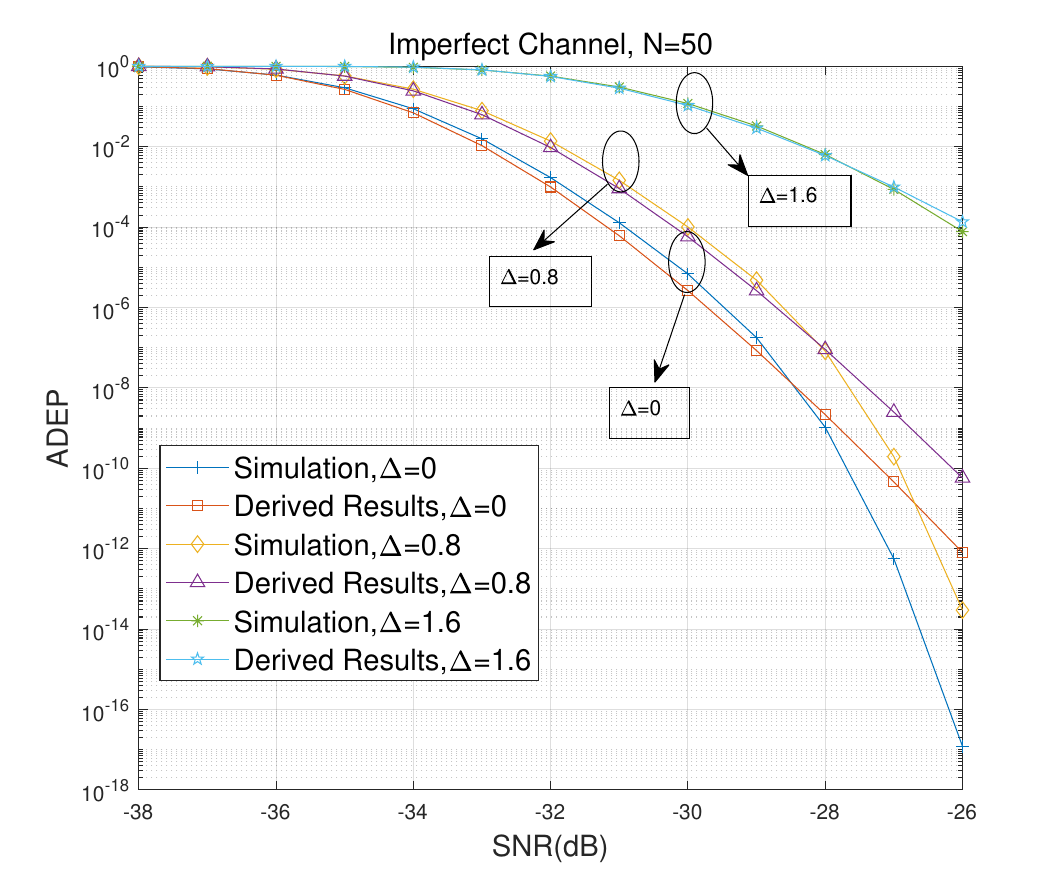}
	}
	\subfigure[$N=100$]{
		\includegraphics[width=0.46\textwidth]{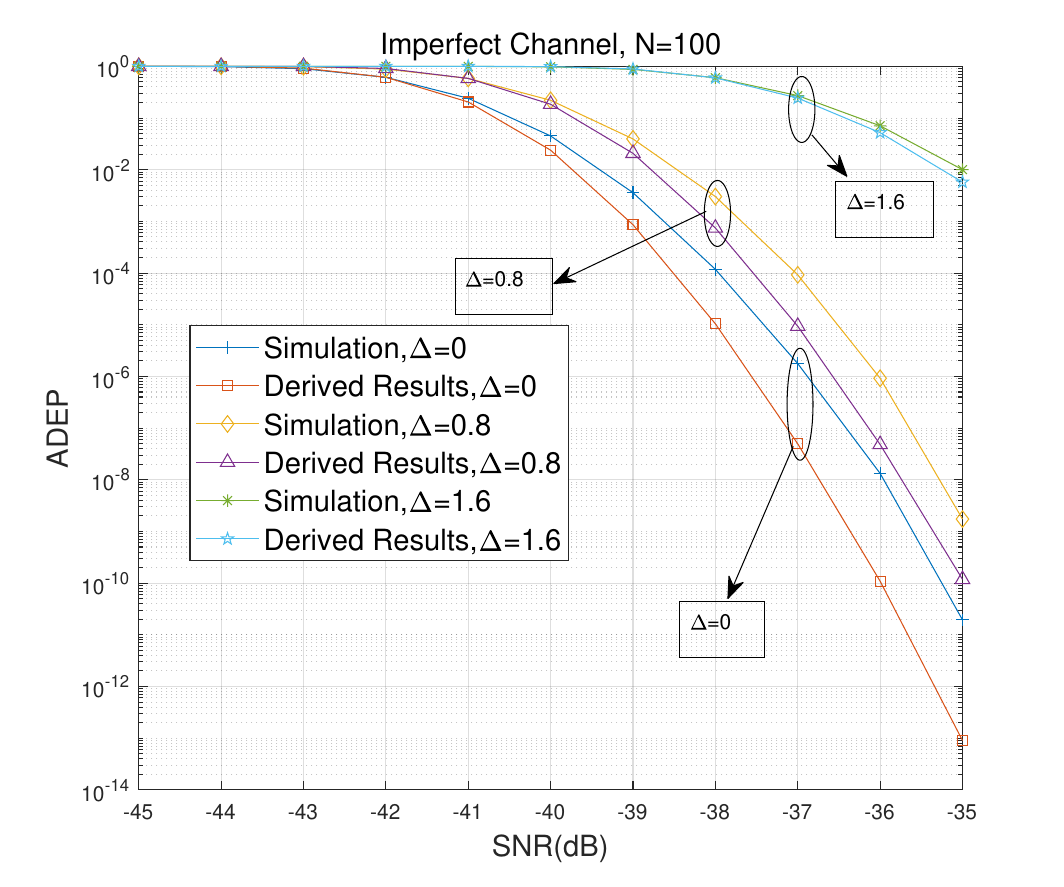}
	}
	\caption{ADEP versus SNR  for the case  of imperfect phase alignment.
	} \label{fig8}
\end{figure}

In this subsection, we study the case when there is imperfect phase alignment. The different values of phase alignment error $\Delta$ are shown in the simulation figures.

In Fig.~\ref{fig7}, the ADR is shown as a function of SNR for two different values of $N$. Three different values of phase alignment errors are considered: $\Delta=0, 1.2, 2$. \textcolor[rgb]{0.00,0.50,1.00}{By comparing Fig.~\ref{fig7} for the case of $\Delta=0$ with Fig.~\ref{fig1}, we can find that the ADR achieved in both cases is the same, which implies the accuracy of our derived results. }As expected, the ADR decreases with the increase of the phase alignment error. In addition, for various values of $\Delta$ and $N$, the derived results are consistent with the simulation results, and verifies the accuracy of the derived results. Larger value of $N$ can achieve larger value of ADR. \textcolor[rgb]{0.00,0.50,1.00}{In addition, higher phase alignment error leads to lower ADR, which means	highly precise phase shifts are required to achieve the satisfactory performance.}

In Fig.~\ref{fig8}, we plot the ADEP versus SNR for two different values of $N$. Three different values of phase alignment errors are considered:
$\Delta=0, 0.8, 1.6$. It is observed from Fig.~\ref{fig8} that the \textcolor[rgb]{0.00,0.50,1.00}{ADEP increases with the increase of the phase alignment error. Specifically, when $N=50$ and SNR is -30 dB, the ADEP increases from $10^{-5}$ to $10^{-1}$. This result demonstrates the importance of obtaining the accurate channel state information.} For the case of $N=100$, the ADR achieved by the derived results has almost the same trend as the simulation results although there is some gap between them.

\vspace{-0.3cm}
\subsection{Multiple-IRS Case}

\begin{figure}[htbp]
	\centering
	\subfigure[$N=50$]{
		\includegraphics[width=0.46\textwidth]{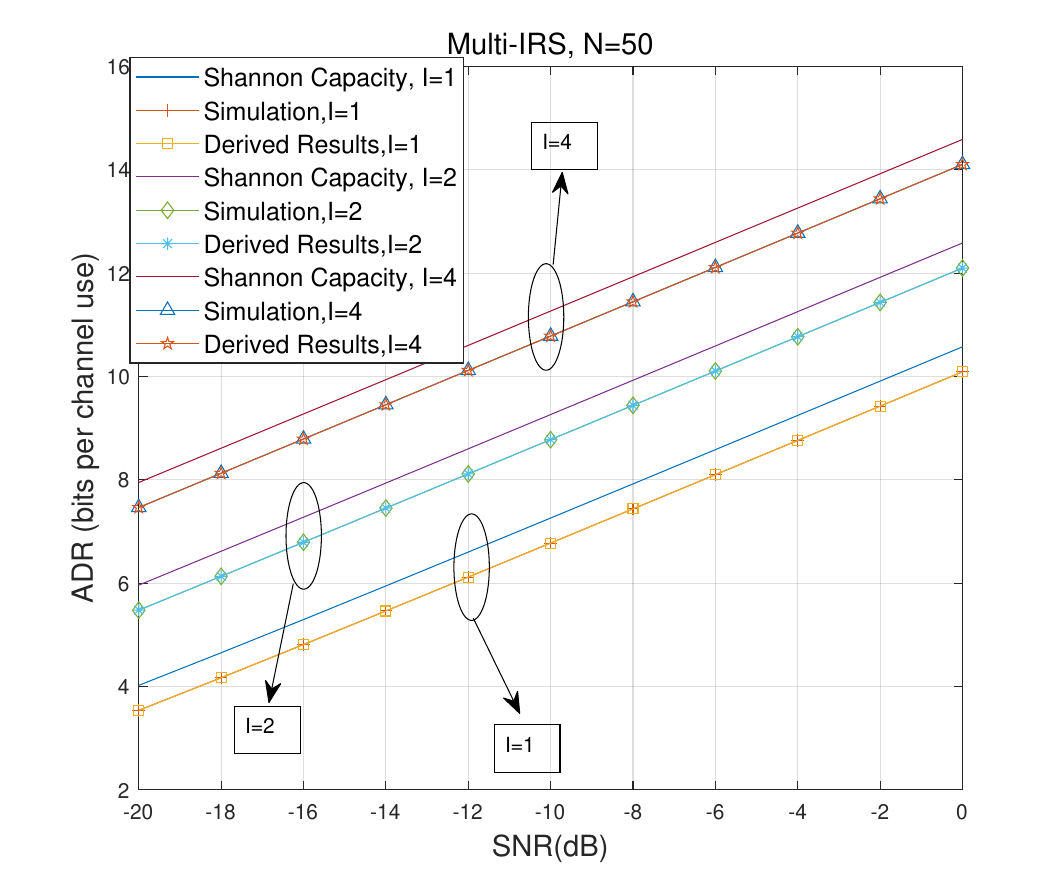}
	}
	\subfigure[$N=100$]{
		\includegraphics[width=0.46\textwidth]{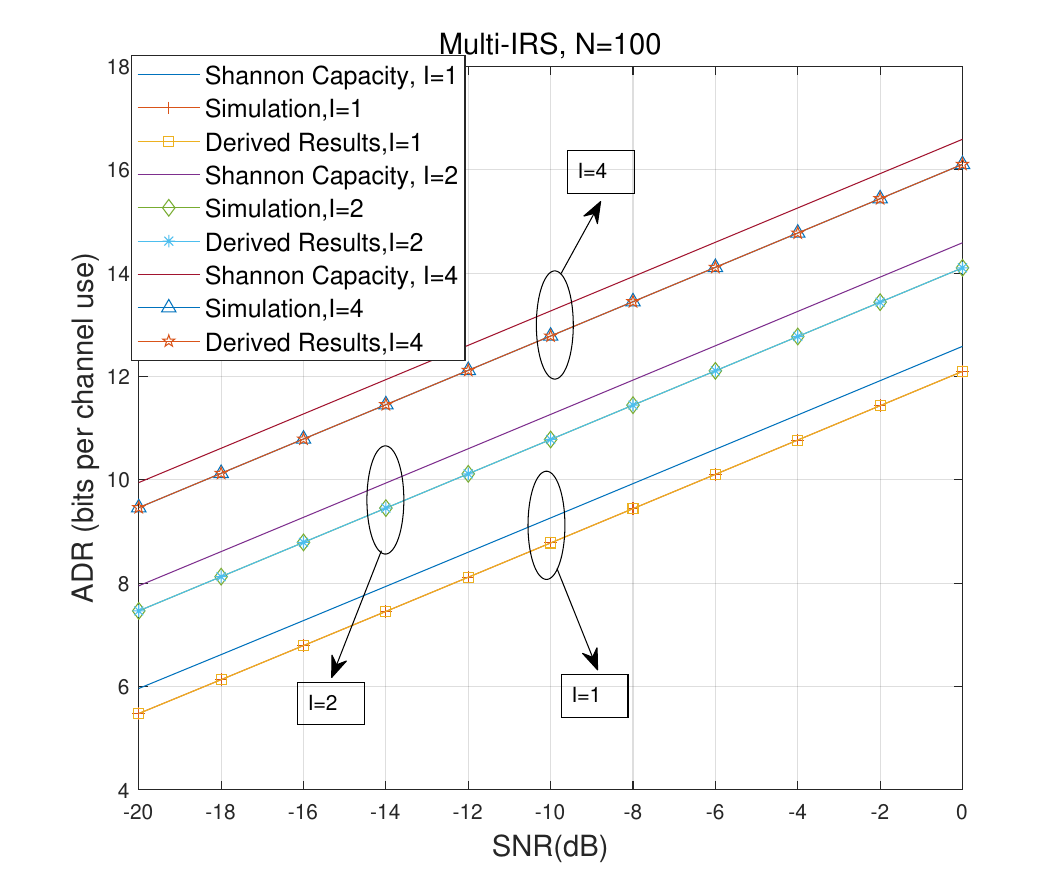}
	}
	\caption{ADR versus SNR  for the case  of multiple IRSs.
	} \label{fig9}
\end{figure}

 \begin{figure}[htbp]
	\centering
	\subfigure[$N=20$]{
		\includegraphics[width=0.46\textwidth]{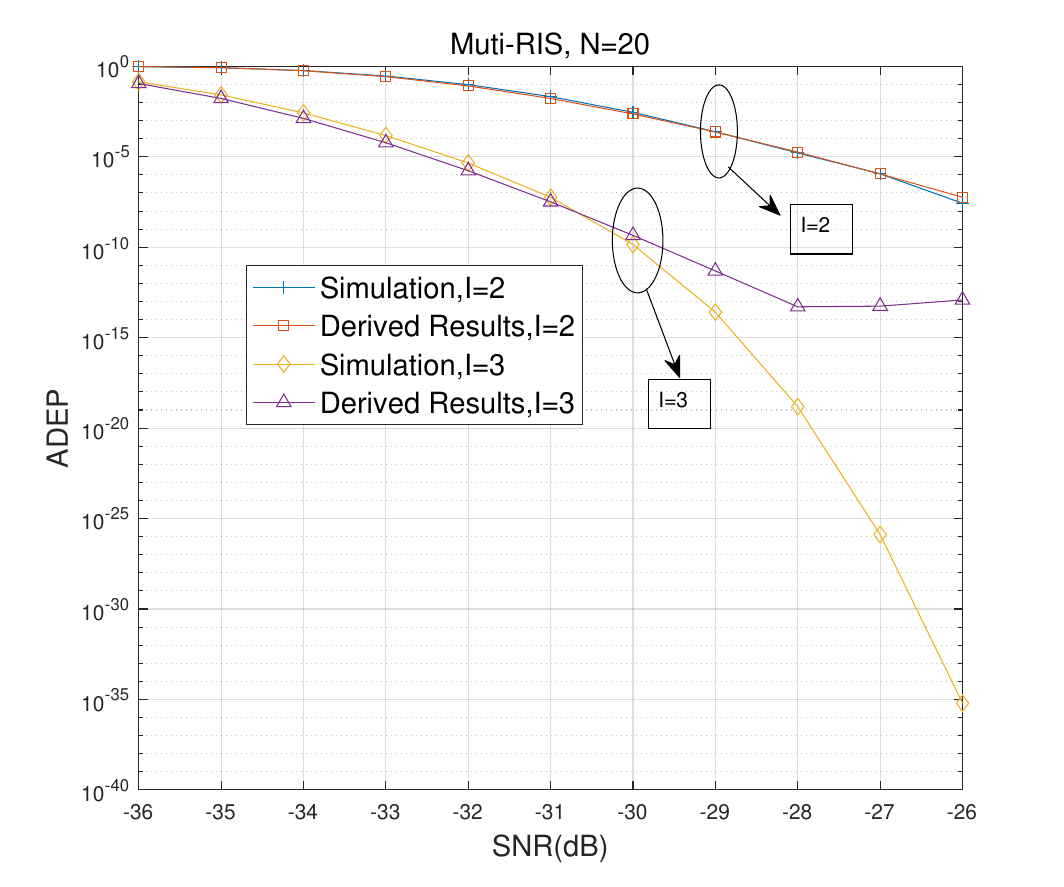}
	}
	\subfigure[$N=40$]{
		\includegraphics[width=0.46\textwidth]{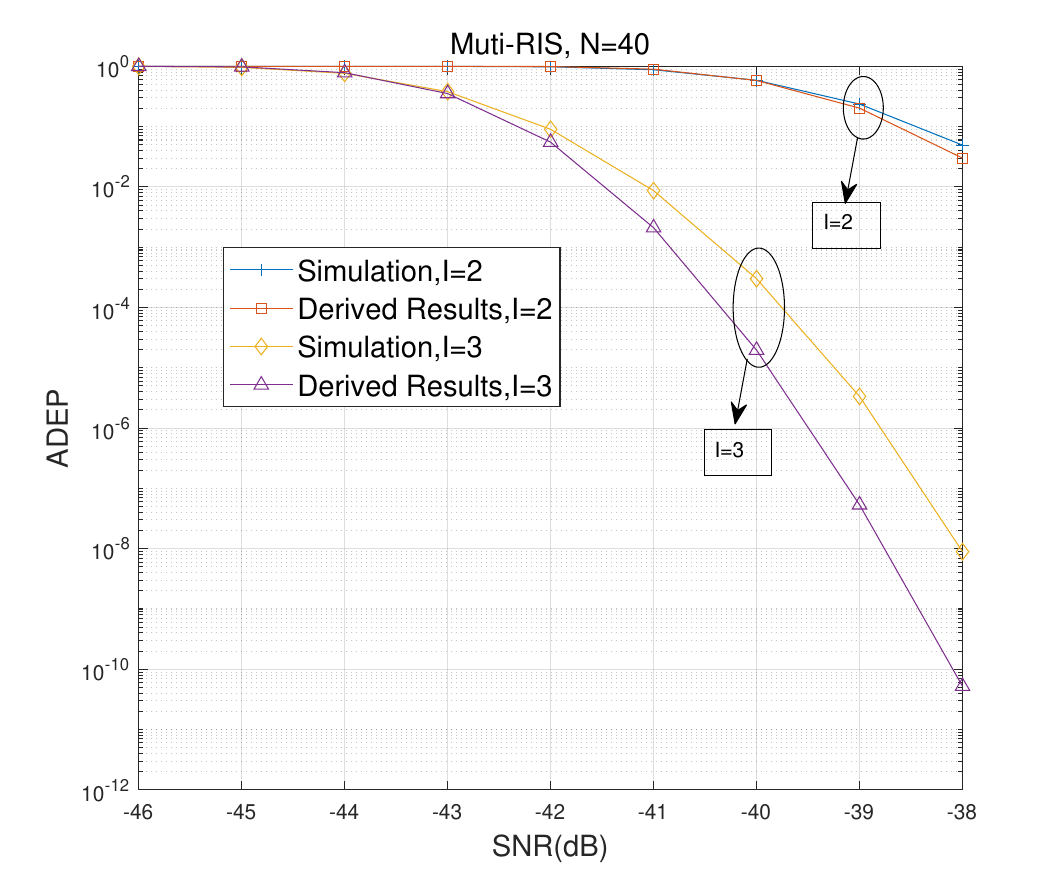}
	}
	\caption{ADEP versus SNR  for the case  of multiple IRSs.
	} \label{fig10}
\end{figure}

In this subsection, we consider the case when there are multiple IRSs in the system. The various numbers of IRSs $I$ are shown in the simulation figures.

In Fig.~\ref{fig9}, we plot the ADR versus SNR for three different values of $I$ and $N$. It is again observed that the ADR increases with SNR, and larger value of $N$ can achieve higher ADR. As expected, more IRSs can achieve higher ADR due to the increased reflecting beamforming gain. \textcolor[rgb]{0.00,0.50,1.00}{Specifically, by comparing Fig.~\ref{fig9}-(a) with Fig. \ref{fig1}-(a), when $I$ increases from 1 to 2  and ${\rm{SNR}}=-12\  \rm{dB}$, the ADR increases from 6 bits per channel use to 8 bits per channel use. Hence, more RISs can increase the ADR performance due to the increased diversity gains.}

In Fig.~\ref{fig10}, the ADEP versus SNR is plotted for various values of $I$ and $N$. \textcolor[rgb]{0.00,0.50,1.00}{We can observe from Fig.~\ref{fig10} that the ADEP decreases with the increase of $I$ due to the increased beamforming gain. In particular, for the case of $N=20$ and ${\rm{SNR}}=-30\  \rm{dB}$, the ADR decreases from $10^{-3}$ to $10^{-10}$ when increasing $I=2$ to $I=3$, which demonstrates the benefits of using more IRSs.} For the case of $N=40$, the derived results have the similar trend as the simulation results although there are some performance differences. This difference is mainly due to the approximation error in the large value of $\gamma$.

\vspace{-0.3cm}
\subsection{Rician Fading Channel}

\begin{figure}[htbp]
	\centering
	\subfigure[$N=50$]{
		\includegraphics[width=0.46\textwidth]{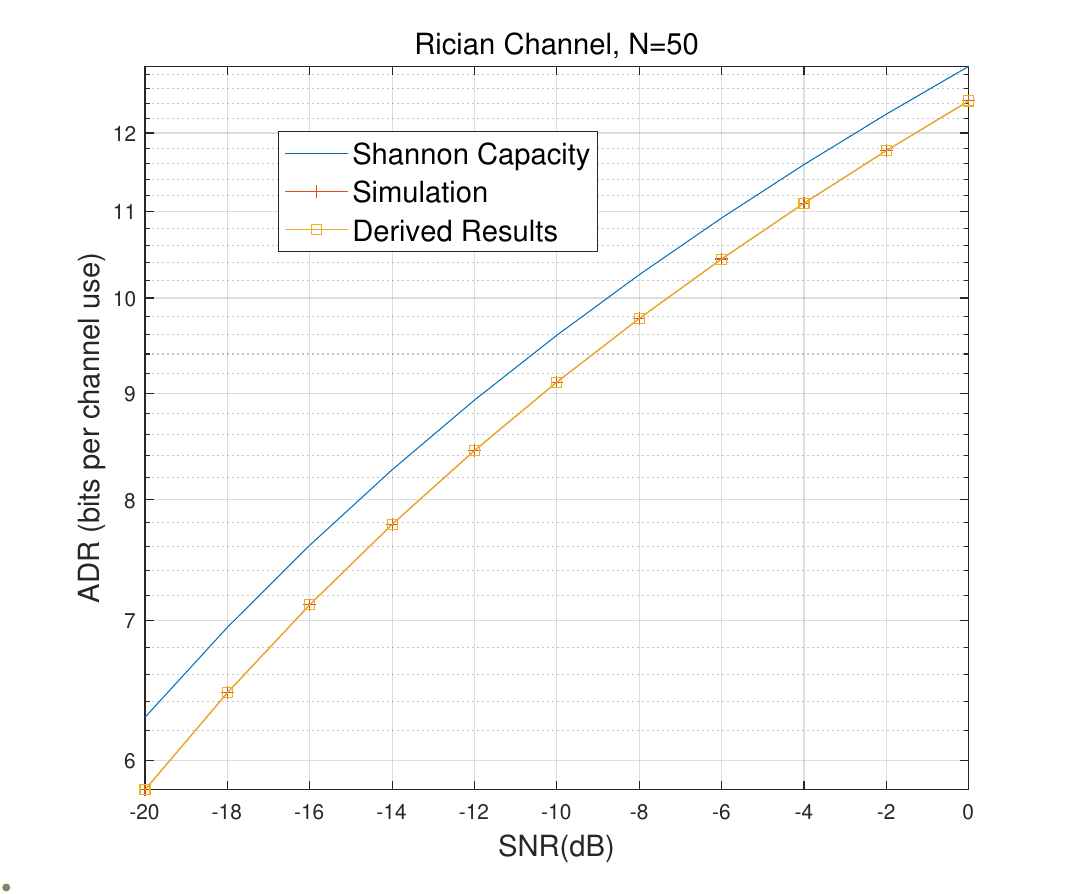}
	}
	\subfigure[$N=100$]{
		\includegraphics[width=0.46\textwidth]{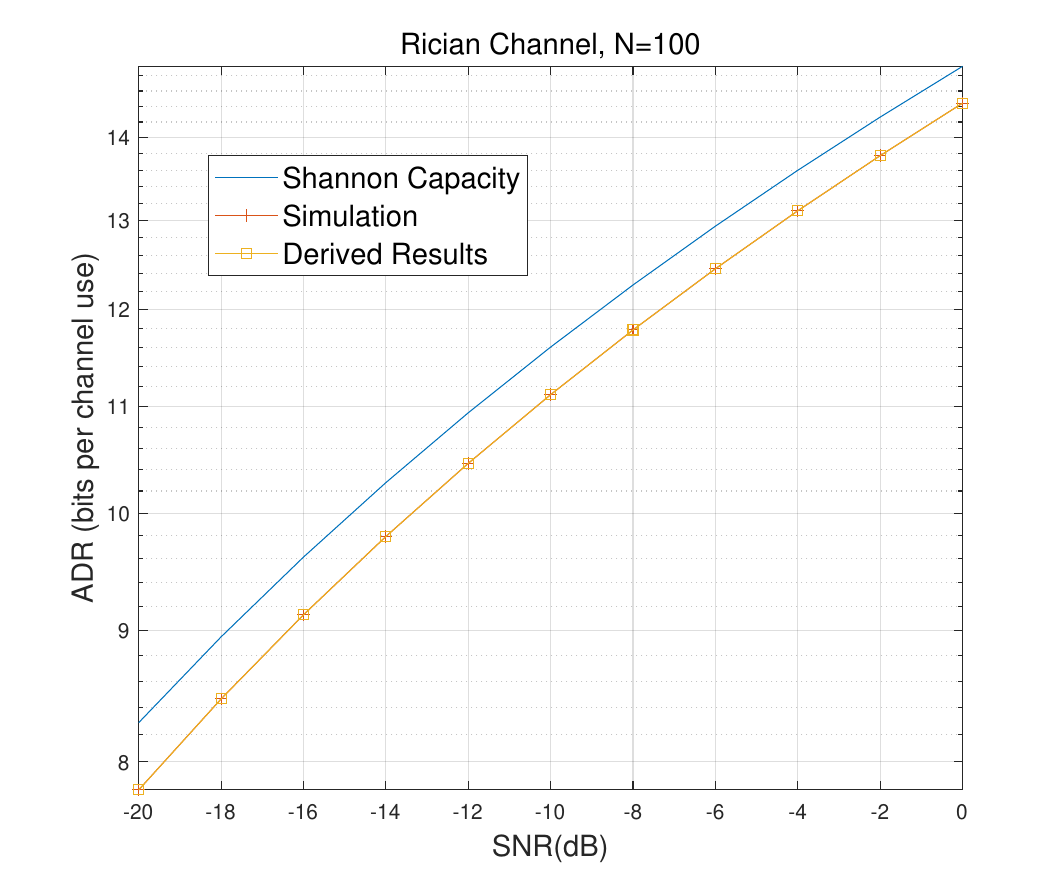}
	}
	\caption{ADR versus SNR  for the case  of Rician channel.
	} \label{fig11}
\end{figure}

 \begin{figure}[htbp]
	\centering
	\subfigure[$N=50$]{
		\includegraphics[width=0.46\textwidth]{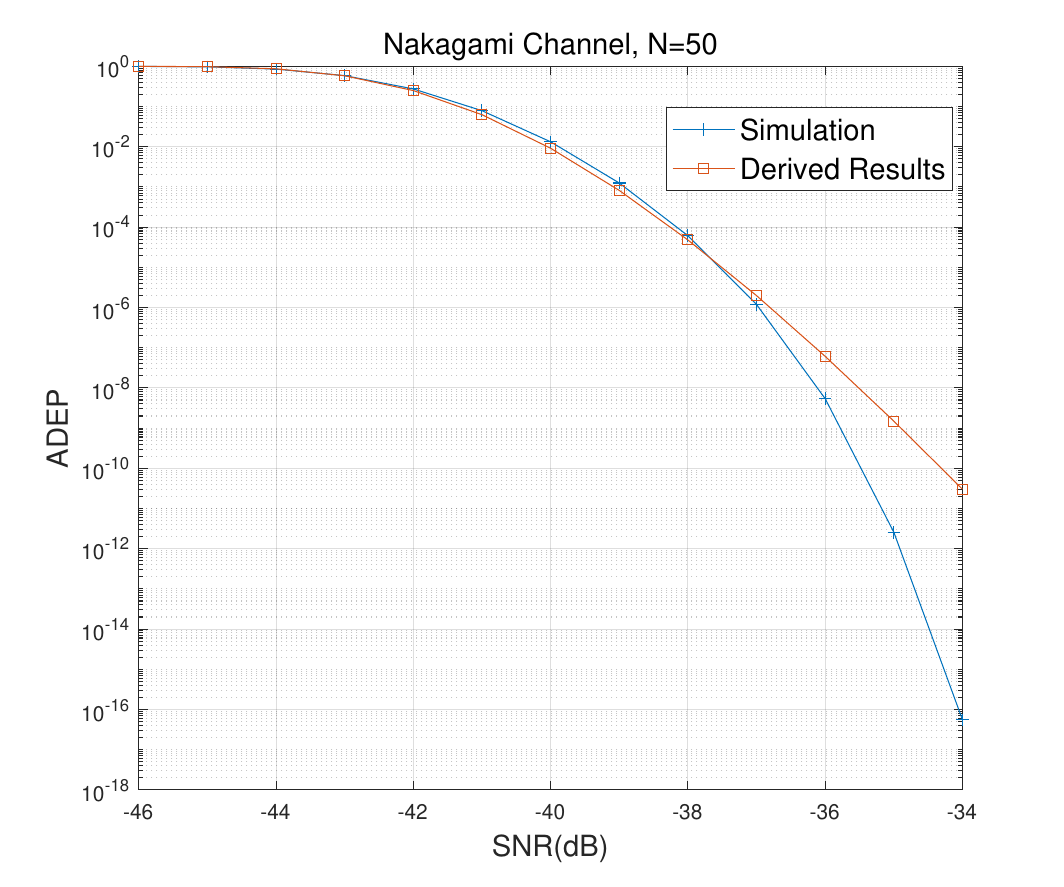}
	}
	\subfigure[$N=100$]{
		\includegraphics[width=0.46\textwidth]{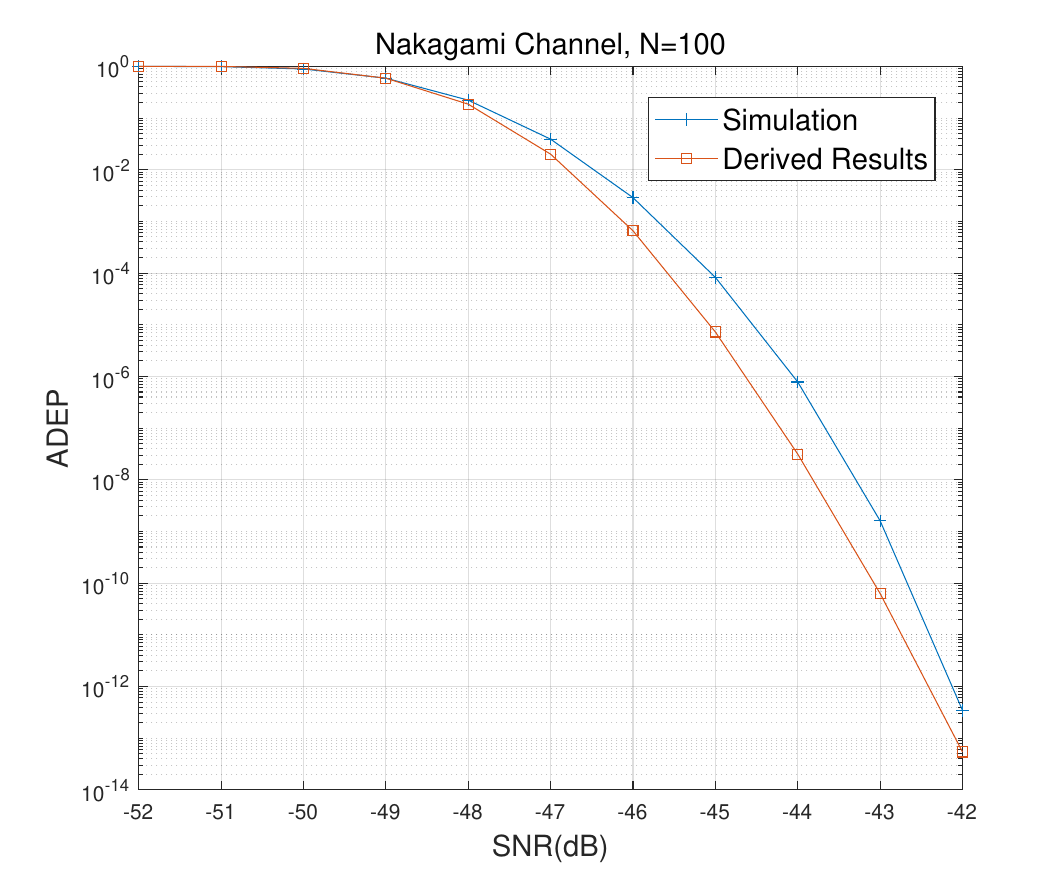}
	}
	\caption{ADEP versus SNR  for the case  of Rician channel.
	} \label{fig12}
\end{figure}

In this subsection, we consider the case of Rician channel model, where the parameters are set as $\alpha_1=\alpha_2=1$, and $\beta_1=\beta_2=0.5$. Then, the parameter that denotes  the ratio of the power contributions by line-of-sight path to the remaining multipaths is $K = 1/8$.

In Fig.~\ref{fig11}, we illustrate the ADR versus SNR for different values of $N$. Again, it is observed that the simulation results coincide with the derived results, which verifies the accuracy of our derived results. \textcolor[rgb]{0.00,0.50,1.00}{For the case of $N=50$ and ${\rm{SNR}}=-10\  \rm{dB}$, 9 bits per channel use can be achieved for the case of Rician channel, while only 6.8 bits per channel use is achieved for the case of Rayleigh channel model. This shows the advantage of using line-of-sight links in IRS-aided communications. }In addition, the benefits of using  more reflecting elements are also demonstrated in Fig.~\ref{fig11}.

In Fig.~\ref{fig12}, the ADEP versus SNR is plotted for two values of $N$. For the case of $N=50$, the derived results are very accurate when ${\rm{SNR}}\le -37\  \rm{dB}$, and the ADEP can be as low as $10^{-6}$, which are enough for some URLLC applications. It is observed there is performance difference between the derived results and the simulation results when ${\rm{SNR}}\ge -38\  \rm{dB}$, which is due the approximation error due to the large value of $\gamma$. The similar phenomenon is observed for the case of $N=100$.

\vspace{-0.3cm}
\subsection{Correlated Channels}
\begin{figure}[htbp]
	\centering
	\subfigure[$N=50$]{
		\includegraphics[width=0.46\textwidth]{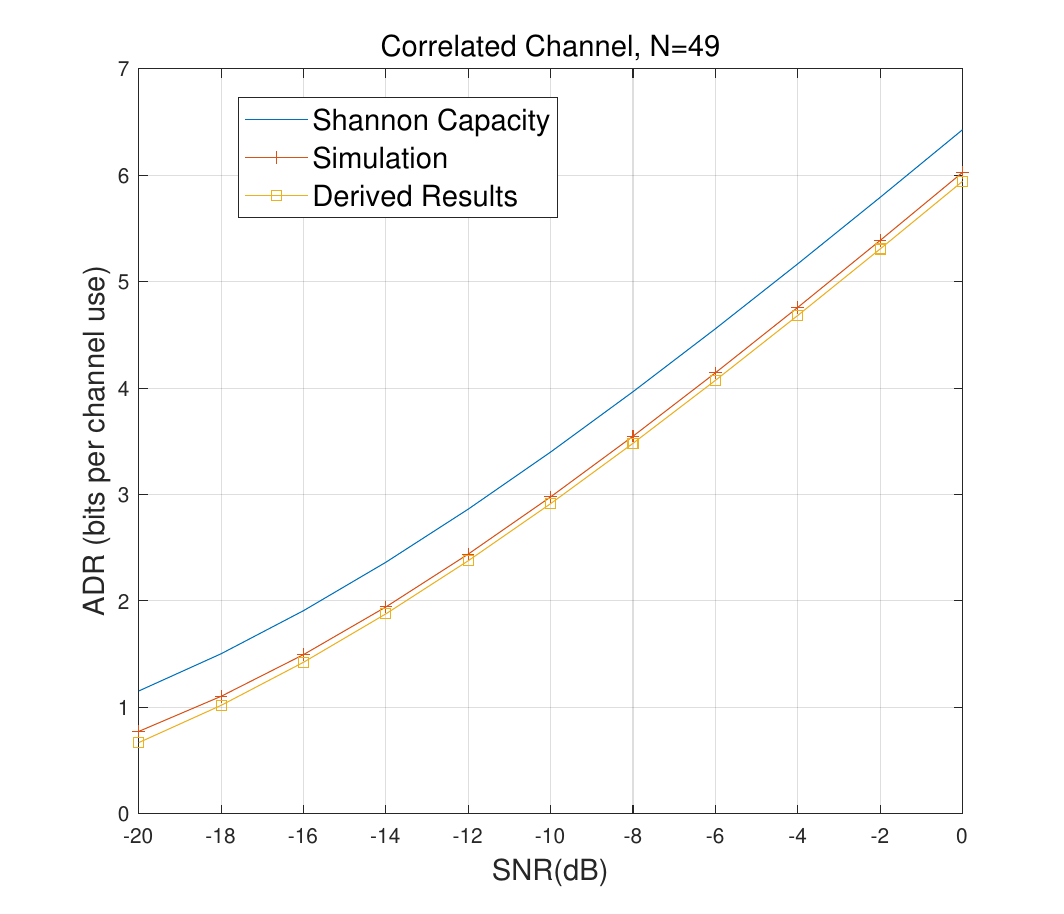}
	}
	\subfigure[$N=100$]{
		\includegraphics[width=0.46\textwidth]{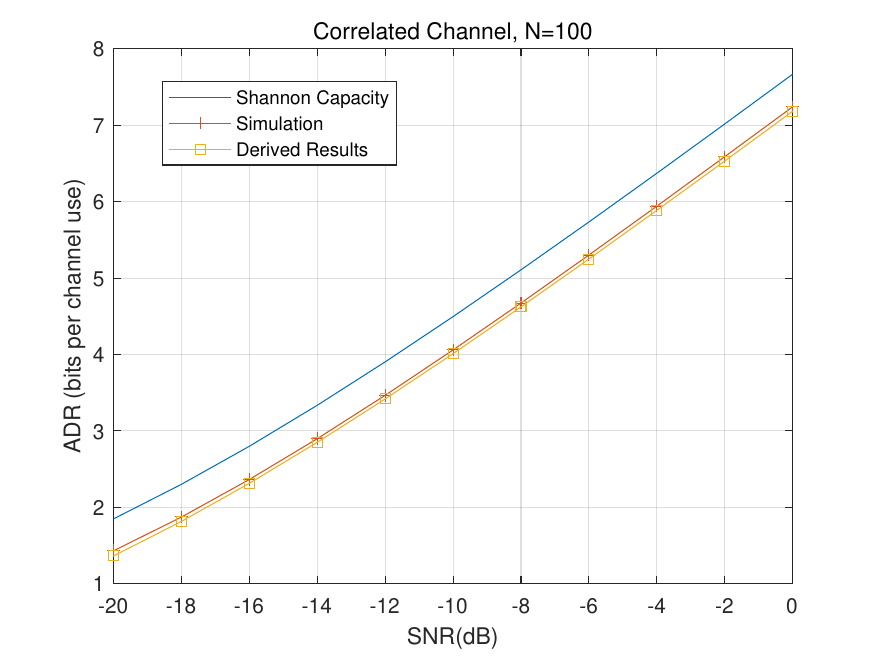}
	}
	\caption{ADR versus SNR  for the case  of correlated channel.
	} \label{fig13}
\end{figure}

 \begin{figure}[htbp]
	\centering
	\subfigure[$N=50$]{
		\includegraphics[width=0.46\textwidth]{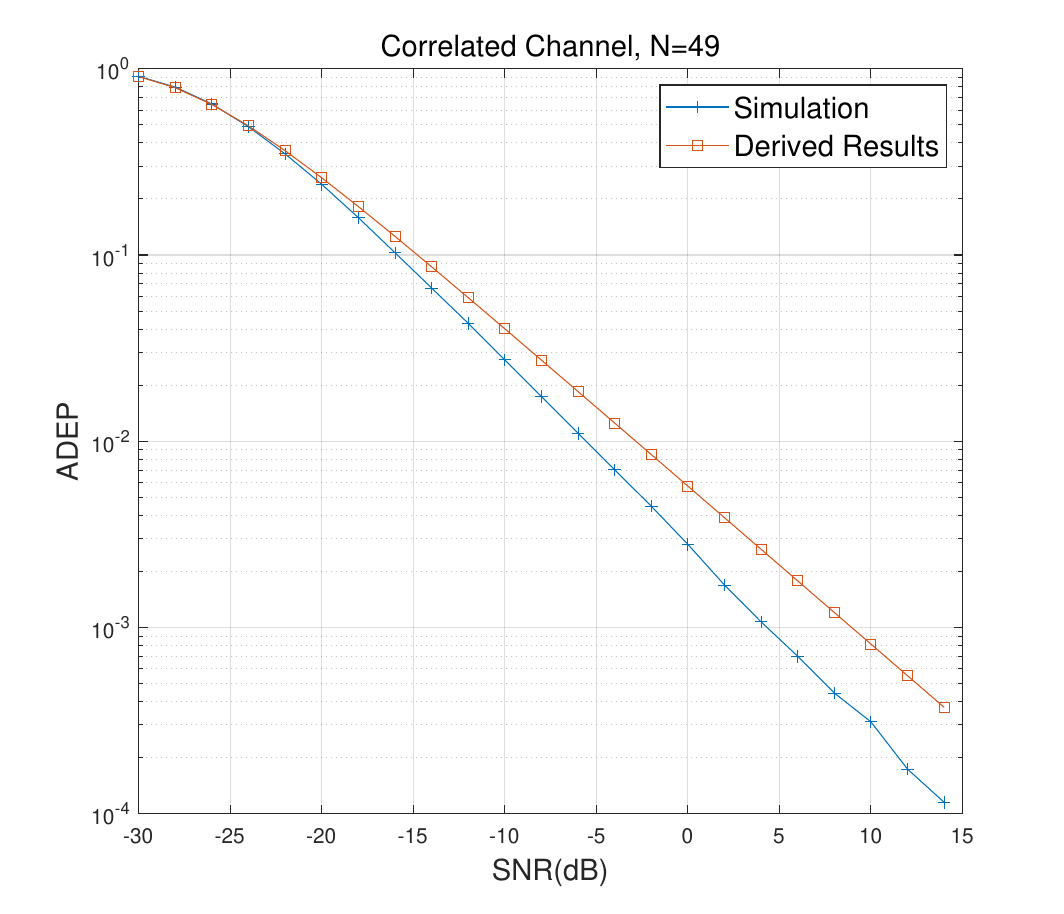}
	}
	\subfigure[$N=100$]{
		\includegraphics[width=0.46\textwidth]{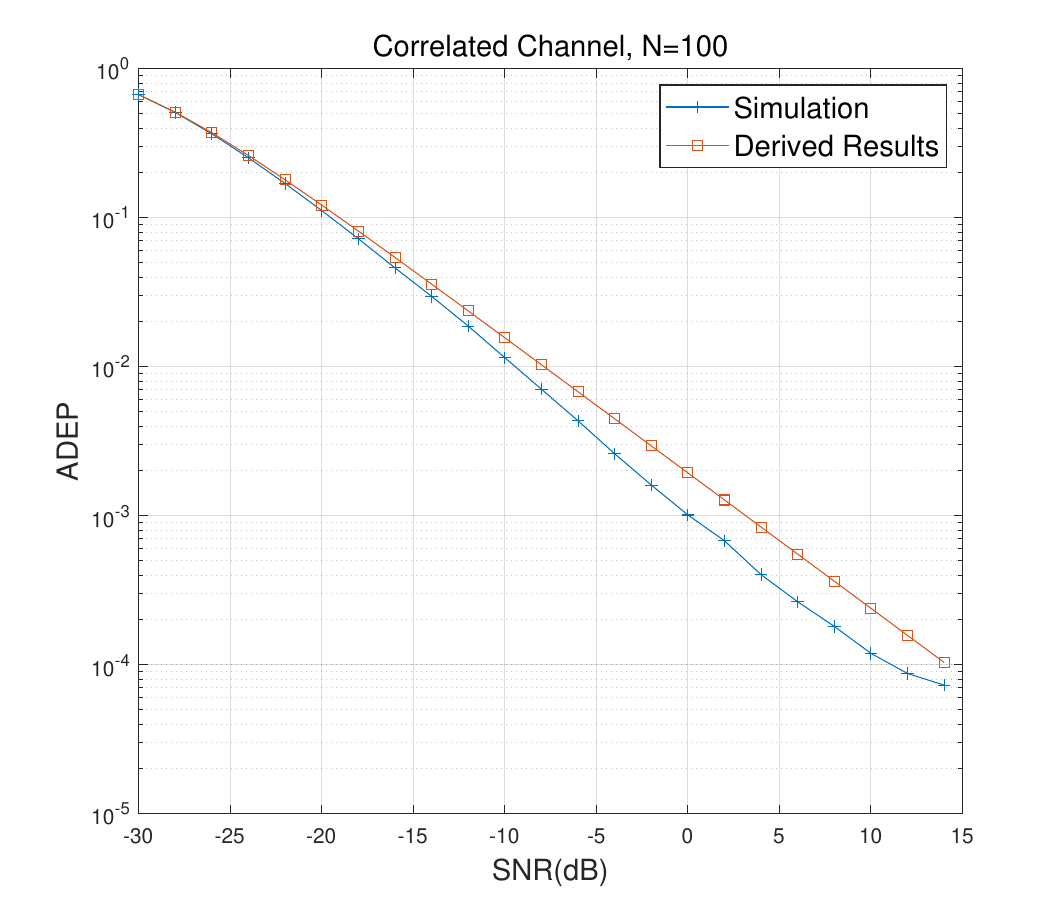}
	}
	\caption{ADEP versus SNR  for the case  of correlated channel.
	} \label{fig14}
\end{figure}

Finally, we consider the case when there is correlation among the reflecting elements at the IRS. The parameters are set as follows: $d_H=d_V=\lambda/4$, and $N_H=N_V$.

In Fig.~\ref{fig13}, we plot the ADR versus the SNR for two different values of $N$. It can be observed from Fig.~\ref{fig13} that the ADR increases with the SNR as expected. \textcolor[rgb]{0.00,0.50,1.00}{However, by comparing Fig.~\ref{fig13} with Fig.~\ref{fig1}, we can find that the performance under the correlated channel is much lower than that of the uncorrelated channel. This implies that the correlation significantly affects the ADR performance, and the element spacing should be large to reduce the coupling and correlation among the elements.}

In Fig.~\ref{fig14}, the ADEP is shown as a function of SNR. \textcolor[rgb]{0.00,0.50,1.00}{It is observed that the ADEP decreases with SNR, and the ADEP achieved in the case of correlated channel is significantly lower than that without correlation. This again demonstrates the importance of enlarging the element spacing distance to eliminate the mutual correlation.}

\vspace{-0.3cm}
\section{Conclusions}\label{conclu}

In this paper, we proposed to deploy an IRS in a FA scenario to support the URLLC. Different from the existing contributions on transmission design, this paper focused on the  analytical analysis by characterizing the performance under the assistance of IRS. To this end, the performance analysis is performed for seven cases:  1) Rayleigh fading channel; 2) With direct channel link; 3)  Nakagami-m fading channel; 4) Imperfect phase alignment; 5) Multiple-IRS case; 6) Rician fading channel; 7) Correlated channels. The approximate/accurate ADR and ADEP expressions were derived in closed-form.   Extensive numerical results were provided to validate the accuracy of our derived results. 

\textcolor[rgb]{0.00,0.50,1.00}{In this work, the phase shifts of the IRS are designed based on the instantaneous CSI, which entails high channel estimation overhead. One promising future direction is to design the phase shifts based on statistical CSI such as angle/location information that varies much slower than the instantaneous CSI. Hence,  channel estimation overhead can be significantly reduced. In addition, only one transmit antenna is assumed to be equipped at the base station. To serve more devices with the same time and frequency resources, multiple-antenna technique is a promising research direction due to the fact that it can provide enough spatial degrees of freedom. However, the performance analysis for this case is more complicated, and would leave it for future work. Furthermore, in smart factories with limited space, the transmission distance between the IRS and the devices may fall into the near-field regime. In this case, the conventional uniform plane wave (UPW)-based channel model may become inaccurate, and the near-field radiation with spherical wavefront may appear. How to analyze the performance in the near filed would be challenging but practical, and we will leave it to our future work.}

\numberwithin{equation}{section}

\begin{appendices}

\section{Proof of Lemma 1}\label{lemma1}
Define ${\xi _n} = \left| {{g_n}} \right|\left| {{h_n}} \right|$. When $|h_n|$ and $|g_n|$ are Rayleigh distributions, we have
\begin{eqnarray}
  \mathbb{E}\left\{ {{\xi _n}} \right\}&=& \frac{\pi }{4}\sqrt {\alpha \beta }, \forall n \label{nhtii}\\
  \mathbb{E}\left\{ {\xi _n^2} \right\} &=&  \mathbb{E}\left\{ {{{\left| {{g_n}} \right|}^2}{{\left| {{h_n}} \right|}^2}} \right\} = \alpha \beta, \forall n\label{kyjokykouy}\\
  \mathbb{E}\left\{ {\xi _n^3} \right\} &=&  \mathbb{E}\left\{ {{{\left| {{g_n}} \right|}^3}{{\left| {{h_n}} \right|}^3}} \right\} = \frac{9}{{16}}\pi{\left( {\alpha \beta } \right)^{\frac{3}{2}}}, \forall n \label{hufehghi} \\
  \mathbb{E}\left\{ {\xi _n^4} \right\}  &=& \mathbb{E}\left\{ {{{\left| {{g_n}} \right|}^4}{{\left| {{h_n}} \right|}^4}} \right\} = 4{\left( {\alpha \beta } \right)^2}, \forall n.\label{jhtyjidei}
\end{eqnarray}
Then, $X$ can be rewritten as $X \buildrel \Delta \over = {\left( {\sum\nolimits_{n = 1}^N {\xi _n} } \right)^2}$. The first moment of the RV $X$ can be calculated by
\begin{equation}\label{grthyth}
  u_X=\mathbb{E}\left\{ X \right\} = \sum\limits_{n = 1}^N {\mathbb{E}\left\{ {\xi _n^2} \right\}}  + \sum\limits_{n = 1}^N {\sum\limits_{m = 1,m \ne n}^N {\mathbb{E}\left\{ {{\xi _n}} \right\}\mathbb{E}\left\{ {{\xi _m}} \right\}} },
\end{equation}
which can be calculated  by using (\ref{nhtii}) and (\ref{kyjokykouy}).

The second moment of $X$ can be calculated as follows:
\begin{eqnarray}
  u_X^{(2)}&=&\mathbb{E}\left\{ {{X^2}} \right\} = \mathbb{E}\left\{ {{{\left( {\sum\limits_{n = 1}^N {{\xi _n}} } \right)}^4}} \right\} = \mathbb{E}\left\{ {{{\left( {\sum\limits_{n = 1}^N {\xi _n^2}  + \sum\limits_{n = 1}^N {\sum\limits_{m = 1,m \ne n}^N {{\xi _n}} } {\xi _m}} \right)}^2}} \right\} \\
     &=& \sum\limits_{n = 1}^N {\mathbb{E}\left\{ {\xi _n^4} \right\}}  + 4\sum\limits_{n = 1}^{N } {\sum\limits_{m =  1,m \ne n}^N {\mathbb{E}\left\{ {\xi _n^3{\xi _m}} \right\}} }  + 3\sum\limits_{n = 1}^{N} {\sum\limits_{m =  1,m\ne n}^N {\mathbb{E}\left\{ {\xi _n^2\xi _m^2} \right\}} } \nonumber  \\
     && + 6\sum\limits_{{n_1} = 1}^{N} {\sum\limits_{{n_2} = 1, n_2\ne n_1}^{N} {\sum\limits_{{n_3} =1, n_3\ne n_2, n_3\ne n_1}^N {\mathbb{E}\left\{ {\xi _{{n_1}}^2{\xi _{{n_2}}}{\xi _{{n_3}}}} \right\}} } } \nonumber \\
     &&+ \sum\limits_{{n_1} = 1}^N {\sum\limits_{{n_2} \ne {n_1}} {\sum\limits_{{n_3} \ne {n_2},{n_3} \ne {n_1}} {\sum\limits_{{n_4} \ne {n_1},{n_2},{n_3}} {\mathbb{E}\left\{ {{\xi _{{n_1}}}{\xi _{{n_2}}}{\xi _{{n_3}}}{\xi _{{n_4}}}} \right\}} } } } .\label{jroigjt}
\end{eqnarray}
 By using the independence between $\xi_n$ and $\xi_m$, and using (\ref{nhtii})-(\ref{jhtyjidei}), we can calculate the second moment of $X$.

\section{Proof of Lemma 2}\label{lemma2}	
By using the approximate PDF of $\gamma $ in (\ref{erdec}),	${\bar R}$ can be approximated as
\begin{eqnarray}
  {\bar R} &=& \int_0^\infty  {\left( {{{\log }_2}\left( {1 + x} \right) - \frac{{{Q^{ - 1}}\left( \varepsilon  \right)}}{{\ln 2}}\sqrt {\frac{{V(x)}}{M}} } \right){f_\gamma }\left( x \right)dx} \nonumber \\
   &=&\!\!\! \frac{1}{{{\theta ^k}\Gamma (k)}}\underbrace {\int_0^\infty  {{\gamma ^{k - 1}}} {e^{ - \frac{\gamma }{\theta }}}{{\log }_2}(1+\gamma)d\gamma }_{{U_1}} - \frac{{{Q^{ - 1}}\left( \varepsilon  \right)}}{{\log (2){\theta ^k}\Gamma (k)}}\underbrace {\int_0^\infty  {{\gamma ^{k - 1}}} \sqrt {\frac{{1 - \frac{1}{{{{(\gamma  + 1)}^2}}}}}{M}} {e^{ - \frac{\gamma }{\theta }}}d\gamma }_{{U_2}}.\label{g5}
\end{eqnarray}

Then, $U_1$ in (\ref{g5}) can be further simplified as
\begin{equation}\label{g38}
	\begin{aligned}
		U_3& =\frac{\left(-\frac{1}{{\theta}^2}\right)^{-{k}} {\theta}^{-{k}-1}}{\log (2)} \\& \left(\left(-\frac{1}{{\theta}}\right)^{{k}} {\theta}^{{k}} \left(\Gamma ({k}-1) \, _2F_2\left(1,1;2,2-{k};\frac{1}{{\theta}}\right)+{\theta} \Gamma ({k}) (\psi ^{(0)}({k})+\log ({\theta}))\right) \right.\\& \left.+  \pi  {\theta} \csc (\pi  {k}) \left(\Gamma ({k})-\Gamma \left({k},-\frac{1}{{\theta}}\right)\right)\right)
		\\&=\frac{(-1)^{-{k}} {\theta}^{{k}-1}}{\log (2)} \left((-1)^{{k}} \left(\Gamma ({k}-1) \, _2F_2\left(1,1;2,2-{k};\frac{1}{{\theta}}\right)+{\theta} \Gamma ({k}) (\psi ^{(0)}({k})  +\log ({\theta}))\right)  \right.\\& \left. + \pi  {\theta} \csc (\pi  {k}) \left(\Gamma ({k})-\Gamma \left({k},-\frac{1}{{\theta}}\right)\right)\right)
	\end{aligned}
\end{equation}
where ${\psi ^{(0)}}(n)$ is the logarithmic Gamma function \cite{Poly,gradshteyn2014table}.

By using $V(\gamma) = 1-\frac{1}{(\gamma+1)^2} \approx 1$, $U_2$ can be written as
\begin{equation}\label{g7}
	\begin{aligned}
		&U_4 \approx \sqrt{\frac{1}{M}} {\theta}^{{k}} \Gamma ({k})
	\end{aligned}
\end{equation}
Then,  ${\bar R}$ in (\ref{g5}) can be approximated as
  \begin{equation}\label{g77}
	\begin{aligned}
		&  {\bar R} \approx \frac{\Gamma ({k}-1) \, _2F_2\left(1,1;2,2-{k};\frac{1}{{\theta}}\right)}{{\theta} (\log (2) \Gamma ({k}))}-\frac{{{Q^{ - 1}}\left( \varepsilon  \right)}}{\sqrt{M} \log (2)}+\frac{\psi ^{(0)}({k})+\log ({\theta})}{\log (2)} \\&+\frac{\left(\pi  (-1)^{-{k}} \csc (\pi  {k})\right) \left(\Gamma ({k})-\Gamma \left({k},-\frac{1}{{\theta}}\right)\right)}{\log (2) \Gamma ({k})}
	\end{aligned}
\end{equation}
which can be further simplified as (\ref{gd8}).

\section{Proof of Lemma 3}\label{lemma3}	
	
We adopt the linearization technique to approximate ${Q\left( {\ln 2\sqrt {\frac{M}{{V(x)}}} \left( {{{\log }_2}(1 + x) - \frac{D}{M}} \right)} \right)}$ at point $x=x_0$ as:
\begin{equation}\label{fregtr}
	\begin{array}{l}
		Q\left( {\ln 2\sqrt {\frac{M}{{V(x)}}} \left( {{{\log }_2}(1 + x) - \frac{D}{M}} \right)} \right)\\
		\approx U(x) = \left\{ \begin{array}{l}
			1,x \le {x_0} - \frac{1}{{2\mu }},\\
			\frac{1}{2} - \mu \left( {x - {x_0}} \right),x \in \left[ {{x_0} - \frac{1}{{2\mu }},{x_0} + \frac{1}{{2\mu }}} \right],\\
			0,x \ge {x_0} + \frac{1}{{2\mu }},
		\end{array} \right.
	\end{array}
\end{equation}
where $\mu$ is the value of the first-order derivative of Q-function at point $x=x_0$, given by
\begin{equation}\label{dsgftr}
	\mu  =  - {\left. {\frac{{\partial \left( {Q\left( {\ln 2\sqrt {\frac{M}{{V(x)}}} \left( {{{\log }_2}(1 + x) - \frac{D}{M}} \right)} \right)} \right)}}{{\partial x}}} \right|_{x = {x_0}}}= \sqrt {\frac{M}{{2\pi \left( {{2^{\frac{{2D}}{M}}} - 1} \right)}}}.
\end{equation}
In addition, $x_0$ is given by ${x_0} = {2^{\frac{D}{M}}} - 1$.

By inserting (\ref{fregtr}) into (\ref{cdscdsr}), the ADEP   $\bar \varepsilon$ can be approximated as
	 \begin{equation}\label{g15}
		\begin{aligned}
			\bar \varepsilon &    \approx   {F_\gamma }\left( {{x_0} - \frac{1}{{2\mu }}} \right) + \left( {\frac{1}{2} + \mu {x_0}} \right)\left( {{F_\gamma }\left( {{x_0} + \frac{1}{{2\mu }}} \right) - {F_\gamma }\left( {{x_0} - \frac{1}{{2\mu }}} \right)} \right)
			- \mu     \int_{{x_0} - \frac{1}{{2\mu }}}^{{x_0} + \frac{1}{{2\mu }}} {x{f_\gamma }\left( x \right)dx}
			\\&  =
			\frac{{\mu}}{{\theta}^{{k}} \Gamma ({k})}
			\underbrace { \int_{{x_0} - \frac{1}{{2\mu }}}^{{x_0} + \frac{1}{{2\mu }}}	
				{\gamma}^{{k}} e^{-\frac{{\gamma}}{{\theta}}}
				 dx}_{U_6}
			\\&\quad +\left(\mu x_0+\frac{1}{2}\right) \left({F_\gamma }\left(\frac{1}{2 \mu}+x_0\right)
			-{F_\gamma }\left(x_0-\frac{1}{2 \mu}\right)\right)   +   {F_\gamma }\left(x_0-\frac{1}{2 \mu}\right).
		\end{aligned}
	\end{equation}
	
	
	
%
	Then, $U_6$ can be solved as
	\begin{equation}\label{gff17}
		\begin{aligned}
		& U_6 =	\left({x_0}-\frac{1}{2 {\mu}}\right)^{{k}+1} E_{-{k}}\left(\frac{{x_0}-\frac{1}{2 {\mu}}}{{\theta}}\right)-\left(\frac{1}{2 {\mu}}+{x_0}\right)^{{k}+1} E_{-{k}}\left(\frac{{x_0}+\frac{1}{2 {\mu}}}{{\theta}}\right)
		\end{aligned}
	\end{equation}
Then, after some transformation, one can have (\ref{g17}).

\section{Proof of Lemma 4}	\label{Lemma4}

Let us define $\xi_n \buildrel \Delta \over = \left| {{g_n}} \right|\left| {{h_n}} \right|, n=1, \cdots, N$ and $\xi_{N+1}=\left| h_0 \right|$. Then, $Y$ can be rewritten as $Y = {\left( {\sum\nolimits_{n = 1}^{N + 1} {{\xi _n}} } \right)^2}$. For $1 \le n \le N$, the moments of $\xi _n$ are given in (\ref{nhtii})-(\ref{jhtyjidei}). For $n=N+1$, the moments of $\xi _n$ are given by
\begin{equation}\label{kiyjfhy}
 \mathbb{E}\left\{ {{\xi _n}} \right\} = \frac{{\sqrt {\pi \eta } }}{2}, \mathbb{E}\left\{ {\xi _n^2} \right\} = \eta , \mathbb{E}\left\{ {\xi _n^3} \right\} = \frac{3}{4}\sqrt \pi  {\eta ^{\frac{3}{2}}}, \mathbb{E}\left\{ {\xi _n^4} \right\} = 2{\eta ^2}, n=N+1.
\end{equation}
Then, by using the similar methods in Appendix \ref{lemma1}, we have
\begin{equation}\label{gdwedwyth}
  u_Y=\mathbb{E}\left\{ Y \right\} = \sum\limits_{n = 1}^{N+1} {\mathbb{E}\left\{ {\xi _n^2} \right\}}  + \sum\limits_{n = 1}^{N+1} {\sum\limits_{m = 1,m \ne n}^{N+1} {\mathbb{E}\left\{ {{\xi _n}} \right\}\mathbb{E}\left\{ {{\xi _m}} \right\}} },
\end{equation}
and
\begin{eqnarray}
     u_Y^{(2)} &=& \sum\limits_{n = 1}^{N+1} {\mathbb{E}\left\{ {\xi _n^4} \right\}}  + 4\sum\limits_{n = 1}^{N+1} {\sum\limits_{m =  1,m \ne n}^{N+1} {\mathbb{E}\left\{ {\xi _n^3{\xi _m}} \right\}} }  + 3\sum\limits_{n = 1}^{N+1} {\sum\limits_{m =  1,m\ne n}^{N+1} {\mathbb{E}\left\{ {\xi _n^2\xi _m^2} \right\}} }   \\
     &&  + 6\sum\limits_{{n_1} = 1}^{N+1} {\sum\limits_{{n_2} = 1, n_2\ne n_1}^{N+1} {\sum\limits_{{n_3} =1, n_3\ne n_2, n_3\ne n_1}^{N+1} {\mathbb{E}\left\{ {\xi _{{n_1}}^2{\xi _{{n_2}}}{\xi _{{n_3}}}} \right\}} } } \\
     &&+ \sum\limits_{{n_1} = 1}^N {\sum\limits_{{n_2} \ne {n_1}} {\sum\limits_{{n_3} \ne {n_2},{n_3} \ne {n_1}} {\sum\limits_{{n_4} \ne {n_1},{n_2},{n_3}} {\mathbb{E}\left\{ {{\xi _{{n_1}}}{\xi _{{n_2}}}{\xi _{{n_3}}}{\xi _{{n_4}}}} \right\}} } } } \label{hehfoji}
\end{eqnarray}

\section{Proof of Lemma 6}	\label{Lemma6}	

Let us define ${\xi _n} = \left| {{g_n}} \right|\left| {{h_n}} \right|$. Then, $G$ can be rewritten as $G\buildrel \Delta \over = {\left| {\sum\nolimits_{n = 1}^N {{\xi _n}{e^{j{\omega _n}}}} } \right|^2}$. The first moment of the RV $G$ can be calculated by
 \begin{eqnarray}
   {u_G} &=& \mathbb{E}\{ G\}  = \mathbb{E}\left\{ {{{\left| {\sum\limits_{n = 1}^N {{\xi _n}{e^{j{\omega _n}}}} } \right|}^2}} \right\}\nonumber \\
    &=& \mathbb{E}\left\{ {\sum\limits_{n = 1}^N {\xi_n^2}  + \sum\limits_{n = 1}^N {\sum\limits_{m = 1,m \ne n}^N {{\xi_n}{\xi_m}\cos ({\omega _n} - {\omega _m})} } } \right\} \nonumber\\
    &=& \sum\limits_{n = 1}^N {\mathbb{E}\left\{ {\xi_n^2} \right\}}  + \sum\limits_{n = 1}^N {\sum\limits_{m = 1,m \ne n}^N {\mathbb{E}\left\{ {{\xi_n}} \right\}\mathbb{E}\left\{ {{\xi_m}} \right\}\mathbb{E}\left\{ {\cos ({\omega _n} - {\omega _m})} \right\}} }\label{grtdwdwedyth}.
 \end{eqnarray}
where $\mathbb{E}\left\{ {\xi_n^2} \right\}$ and $\mathbb{E}\left\{ {{\xi_n}} \right\}$  can be calculated  by using (\ref{nhtii}) and (\ref{kyjokykouy}). By using some simple calculations, we have $\mathbb{E}\left\{ {\cos ({\omega _n} - {\omega _m})} \right\}={\left( {\frac{{\sin \Delta }}{\Delta }} \right)^2}, \forall m \ne n$.

The second moment of $G$ can be calculated as follows:
\begin{eqnarray}
  u_G^{(2)}&=& \mathbb{E}\left\{ {{{\left| {\sum\limits_{n = 1}^N {{\xi_n}{e^{j{\omega _n}}}} } \right|}^4}} \right\} = \mathbb{E}\left\{ {{{\left( {\sum\limits_{n = 1}^N {\xi_n^2}  + \sum\limits_{n = 1}^N {\sum\limits_{m = 1,m \ne n}^N {{\xi_n}{\xi_m}\cos ({\omega _n} - {\omega _m})} } } \right)}^2}} \right\} \nonumber \\
     &=&{\sum\limits_{n = 1}^N {\mathbb{E}\left\{ {\xi_n^4} \right\}}  + 2\sum\limits_{n = 1}^N {\sum\limits_{m \ne n} {\mathbb{E}\left\{ {\xi_n^2\xi_m^2{{\cos }^2}\left( {{\omega _n} - {\omega _m}} \right)} \right\}} } }\nonumber  \\
     &&  + \sum\limits_{n = 1}^N {\sum\limits_{m \ne n} {\mathbb{E}\left\{ {\xi_n^2\xi_m^2} \right\}} }  + 4\sum\limits_{n = 1}^N {\sum\limits_{m \ne n} {\mathbb{E}\left\{ {\xi_n^3{\xi_m}\cos \left( {{\omega _n} - {\omega _m}} \right)} \right\}} }  \nonumber \\
     &&  + 2\sum\limits_{{n_1} = 1}^N {\sum\limits_{{n_2} \ne {n_1}} {\sum\limits_{{n_3} \ne {n_1},{n_2}} {\mathbb{E}\left\{ {\xi_{{n_1}}^2{\xi_{{n_2}}}{\xi_{{n_3}}}\cos \left( {{\omega _{{n_2}}} - {\omega _{{n_3}}}} \right)} \right\}} } }   \nonumber \\
     && { + 4\sum\limits_{{n_1} = 1}^N {\sum\limits_{{n_2} \ne {n_1}} {\sum\limits_{{n_3} \ne {n_1},{n_2}} {\mathbb{E}\left\{ {\xi_{{n_1}}^2{\xi_{{n_2}}}{\xi_{{n_3}}}\cos \left( {{\omega _{{n_1}}} - {\omega _{{n_2}}}} \right)\cos \left( {{\omega _{{n_1}}} - {\omega _{{n_3}}}} \right)} \right\}} } } } \nonumber\\
     &&{ + \sum\limits_{{n_1} = 1}^N {\sum\limits_{{n_2} \ne {n_1}} {\sum\limits_{{n_3} \ne {n_1},{n_2}} {\sum\limits_{{n_4} \ne {n_1},{n_2},{n_3}} {\mathbb{E}\left\{ {{\xi_{{n_1}}}{\xi_{{n_2}}}{\xi_{{n_3}}}{\xi_{{n_4}}}\cos \left( {{\omega _{{n_1}}} - {\omega _{{n_2}}}} \right)\cos \left( {{\omega _{{n_3}}} - {\omega _{{n_4}}}} \right)} \right\}} } } } }. \label{dewdwegjt}
\end{eqnarray}
By using some simple calculations, we have
\begin{eqnarray}
  \mathbb{E}\left\{ {{{\cos }^2}\left( {{\omega _n} - {\omega _m}} \right)} \right\} &=& \frac{1}{2} + \frac{1}{2}{\left( {\frac{{\sin \left( {2\Delta } \right)}}{{2\Delta }}} \right)^2} \\
  \mathbb{E}\left\{ {\cos \left( {{\omega _{{n_1}}} - {\omega _{{n_2}}}} \right)\cos \left( {{\omega _{{n_1}}} - {\omega _{{n_3}}}} \right)} \right\} &=& \frac{{\sin (\Delta )\cos \left( \Delta  \right) + 4\Delta {{\sin }^2}(\Delta ) - \sin (\Delta )\cos (3\Delta )}}{{8{\Delta ^3}}}.
\end{eqnarray}
 Then, by using (\ref{nhtii})-(\ref{jhtyjidei}) and the above equalities, we can calculate the second moment of $G$.

\section{Proof of Lemma 7}	\label{Lemma7}
For each $i$, we have
\begin{eqnarray}
  \mathbb{E}\left\{ {{\xi _{i,n}}} \right\}&=& \frac{\pi }{4}\sqrt {\alpha_i \beta_i }, \forall n \label{nhcsdcsii}\\
  \mathbb{E}\left\{ {\xi _{i,n}^2} \right\} &=&  \mathbb{E}\left\{ {{{\left| {{g_{i,n}}} \right|}^2}{{\left| {{h_n}} \right|}^2}} \right\} = \alpha_i \beta_i, \forall n\label{kyjsdcsykouy}\\
  \mathbb{E}\left\{ {\xi _{i,n}^3} \right\} &=&  \mathbb{E}\left\{ {{{\left| {{g_{i,n}}} \right|}^3}{{\left| {{h_{i,n}}} \right|}^3}} \right\} = \frac{9}{{16}}\pi{\left( {\alpha_i \beta_i } \right)^{\frac{3}{2}}}, \forall n \label{hufecdhi} \\
  \mathbb{E}\left\{ {\xi _{i,n}^4} \right\}  &=& \mathbb{E}\left\{ {{{\left| {{g_{i,n}}} \right|}^4}{{\left| {{h_{i,n}}} \right|}^4}} \right\} = 4{\left( {\alpha_i \beta_i } \right)^2}, \forall n.\label{jhtscsidei}
\end{eqnarray}

Based on (\ref{nhcsdcsii})-(\ref{jhtscsidei}), we can now calculate the moments of $\xi _i$ as follows:
\begin{eqnarray}
  \mathbb{E}\left\{ {{\xi _i}} \right\} &=& \frac{\pi }{4}N\sqrt {{\alpha _i}{\beta _i}}\label{tjgrohg}  \\
  \mathbb{E}\left\{ {\xi _i^2} \right\}&=& N{\alpha _i}{\beta _i}{\rm{ + }}\frac{{{\pi ^2}}}{{16}}N(N - 1){\alpha _i}{\beta _i} \label{frjeofjrt}\\
  \mathbb{E}\left\{ {\xi _i^3} \right\} &=& N\frac{9}{{16}}\pi {\left( {{\alpha _i}{\beta _i}} \right)^{\frac{3}{2}}} + 3N(N - 1)\frac{\pi }{4}{\left( {{\alpha _i}{\beta _i}} \right)^{\frac{3}{2}}} + N(N - 1)(N - 2)\frac{{{\pi ^3}}}{{64}}{\left( {{\alpha _i}{\beta _i}} \right)^{\frac{3}{2}}}\label{khuyjyu}\\
 \mathbb{E}\left\{ {\xi _i^4} \right\}&=&4N{\left( {{\alpha _i}{\beta _i}} \right)^2} + \frac{9}{{16}}{\pi ^2}N(N - 1){\left( {{\alpha _i}{\beta _i}} \right)^2} + 3N(N - 1){\left( {{\alpha _i}{\beta _i}} \right)^2}\nonumber\\
   &&+ \frac{{3{\pi ^2}}}{8}N(N - 1)(N - 2){\left( {{\alpha _i}{\beta _i}} \right)^2} + \frac{{{\pi ^4}}}{{256}}N(N - 1)(N - 2)(N - 3){\left( {{\alpha _i}{\beta _i}} \right)^2}.\label{gjrtoigjt}
\end{eqnarray}

Then, based on (\ref{tjgrohg})-(\ref{gjrtoigjt}), we can calculate the first and second moments of $U$:
\begin{eqnarray}
 u_{U} &=&\mathbb{E}\left\{ U \right\}= \mathbb{E}\left\{ {{{\left( {\sum\limits_{i = 1}^I {{\xi _i}} } \right)}^2}} \right\} = \sum\limits_{i = 1}^I {\mathbb{E}\left\{ {\xi _i^2} \right\}}  + \sum\limits_{i = 1}^I {\sum\limits_{j \ne i} {\mathbb{E}\left\{ {{\xi _i}{\xi _j}} \right\}} } \label{fjrtogt}\\
  u_{U}^{(2)} &=&\mathbb{E}\left\{ {{U^2}} \right\}= \sum\limits_{i = 1}^I {\mathbb{E}\left\{ {\xi _i^4} \right\}}  + 4\sum\limits_{i = 1}^I {\sum\limits_{j \ne i} {\mathbb{E}\left\{ {\xi _i^3{\xi _j}} \right\}} }  + 3\sum\limits_{i = 1}^I {\sum\limits_{ j \ne i}  {\mathbb{E}\left\{ {\xi _i^2\xi _j^2} \right\}} }\nonumber \\
   &&  + 6\sum\limits_{{i_1} = 1}^I {\sum\limits_{{i_2} \ne {i_1}}  {\sum\limits_{{i_3} \ne {i_1} {i_2}}  {\mathbb{E}\left\{ {\xi _{{i_1}}^2{\xi _{{i_2}}}{\xi _{{i_3}}}} \right\}} } } \nonumber \\
   && + \sum\limits_{{i_1} = 1}^I {\sum\limits_{{i_2} \ne {i_1}} {\sum\limits_{{i_3} \ne {i_1}, {i_2}} {\sum\limits_{{i_4} \ne {i_1},{i_2},{i_3}} {\mathbb{E}\left\{ {{\xi _{{i_1}}}{\xi _{{i_2}}}{\xi _{{i_3}}}{\xi _{{i_4}}}} \right\}} } } }.\label{vjhodn}
\end{eqnarray}

\section{Proof of Lemma 9}	\label{Lemma9}	

The first moment of $\gamma$ is given by
\begin{equation}\label{HJYOTIJIO}
u_\gamma=\mathbb{E}\{ \gamma \}  = \mathbb{E}\left\{ {\rho {{\left| {{{\bf{g}}^{\rm{H}}}{\bm{\Phi}} {\bf{h}}} \right|}^2}} \right\} = \mathbb{E}\left\{ {\rho {{\bf{g}}^{\rm{H}}}{\bm{\Phi}} {\bf{h}}{{\bf{h}}^{\rm{H}}}{{\bm{\Phi}} ^{\rm{H}}}{\bf{g}}} \right\} = \rho \alpha \beta {\rm{tr}}\left( {{\bf{R}}{{\bm{\Phi}} ^{\rm{H}}}{\bf{R}}{\bm{\Phi}} } \right).
\end{equation}

The second moment of $\gamma$ is given by
\begin{equation}\label{yu7i89o}
u_{\gamma}^{(2)}=\mathbb{E}\{ {\gamma ^2}\}  = {\rho ^2}\mathbb{E}\left\{ {{{\left\| {\sqrt \beta  {{\bf{R}}^{1/2}}{\bm{\Phi}} {\bf{h}}} \right\|}^4}{{\left| {\frac{{{{\bf{g}}^{\rm{H}}}{\bm{\Phi}} {\bf{h}}}}{{\left\| {\sqrt \beta  {{\bf{R}}^{1/2}}{\bm{\Phi}} {\bf{h}}} \right\|}}\frac{{{{\bf{h}}^{\rm{H}}}{{\bm{\Phi}} ^{\rm{H}}}{\bf{g}}}}{{\left\| {\sqrt \beta  {{\bf{R}}^{1/2}}{\bm{\Phi}} {\bf{h}}} \right\|}}} \right|}^2}} \right\}.
\end{equation}
Define $\kappa  = {{{{\bf{g}}^{\rm{H}}}{\bm{\Phi}} {\bf{h}}} \mathord{\left/
 {\vphantom {{{{\bf{g}}^{\rm{H}}}{\bm{\Phi}} {\bf{h}}} {\left\| {\sqrt \beta  {{\bf{R}}^{1/2}}{\bm{\Phi}} {\bf{h}}} \right\|}}} \right.
 \kern-\nulldelimiterspace} {\left\| {\sqrt \beta  {{\bf{R}}^{1/2}}{\bm{\Phi}} {\bf{h}}} \right\|}}$, which is a circularly symmetric Gaussian variable when conditioning on ${\bf{h}}$.  Due to the normalization operation, we have  $\kappa \sim {\cal{CN}}(0, 1)$. In addition, $\kappa$ is also independent of ${\sqrt \beta  {{\bf{R}}^{1/2}}{\bm{\Phi}} {\bf{h}}}$. Hence, we have
 \begin{eqnarray}
   u_\gamma ^{(2)} &=& {\rho ^2}\mathbb{E}\left\{ {{{\left\| {\sqrt \beta  {{\bf{R}}^{1/2}}{\bm{\Phi}} {\bf{h}}} \right\|}^4}{{\left| \kappa  \right|}^4}} \right\} = {\rho ^2}\mathbb{E}\left\{ {{{\left\| {\sqrt \beta  {{\bf{R}}^{1/2}}{\bm{\Phi}} {\bf{h}}} \right\|}^4}} \right\}\mathbb{E}\left\{ {{{\left| \kappa  \right|}^4}} \right\} \nonumber\\
    &=& 2{\rho ^2}{\alpha ^2}{\beta ^2}\left( {{{\left| {{\rm{tr}}\left( {{\bf{R}}{{\bm{\Phi}} ^{\rm{H}}}{\bf{R}}{\bm{\Phi}} } \right)} \right|}^2} + {\rm{tr}}\left( {{{\left( {{\bf{R}}{{\bm{\Phi}} ^{\rm{H}}}{\bf{R}}{\bm{\Phi}} } \right)}^2}} \right)} \right)\label{fehfir}
 \end{eqnarray}
where the last equality is obtained by using \cite{tulino2004random}.
\end{appendices}	
	
%
%
%
%
%
%
%

\
\





\bibliographystyle{IEEEtran}
\bibliography{myre}

\begin{thebibliography}{10}
\providecommand{\url}[1]{#1}
\csname url@samestyle\endcsname
\providecommand{\newblock}{\relax}
\providecommand{\bibinfo}[2]{#2}
\providecommand{\BIBentrySTDinterwordspacing}{\spaceskip=0pt\relax}
\providecommand{\BIBentryALTinterwordstretchfactor}{4}
\providecommand{\BIBentryALTinterwordspacing}{\spaceskip=\fontdimen2\font plus
\BIBentryALTinterwordstretchfactor\fontdimen3\font minus
  \fontdimen4\font\relax}
\providecommand{\BIBforeignlanguage}[2]{{%
\expandafter\ifx\csname l@#1\endcsname\relax
\typeout{** WARNING: IEEEtran.bst: No hyphenation pattern has been}%
\typeout{** loaded for the language `#1'. Using the pattern for}%
\typeout{** the default language instead.}%
\else
\language=\csname l@#1\endcsname
\fi
#2}}
\providecommand{\BIBdecl}{\relax}
\BIBdecl

\bibitem{schwab2017fourth}
K.~Schwab, \emph{The fourth industrial revolution}.\hskip 1em plus 0.5em minus
  0.4em\relax Currency, 2017.

\bibitem{zhibo2017}
Z.~{Pang}, M.~{Luvisotto}, and D.~{Dzung}, ``Wireless high-performance
  communications: The challenges and opportunities of a new target,''
  \emph{IEEE Industrial Electronics Magazine}, vol.~11, no.~3, pp. 20--25,
  2017.

\bibitem{wu2019towards}
Q.~Wu and R.~Zhang, ``Towards smart and reconfigurable environment: Intelligent
  reflecting surface aided wireless network,'' \emph{IEEE Communications
  Magazine}, vol.~58, no.~1, pp. 106--112, 2019.

\bibitem{Basar2019}
E.~{Basar}, M.~{Di Renzo}, J.~{De Rosny}, M.~{Debbah}, M.~{Alouini}, and
  R.~{Zhang}, ``Wireless communications through reconfigurable intelligent
  surfaces,'' \emph{IEEE Access}, vol.~7, pp. 116\,753--116\,773, 2019.

\bibitem{pan2020reconfigurable}
C.~Pan, H.~Ren, K.~Wang, J.~F. Kolb, M.~Elkashlan, M.~Chen, M.~Di~Renzo,
  Y.~Hao, J.~Wang, A.~L. Swindlehurst, X.~You, and L.~Hanzo, ``Reconfigurable
  intelligent surfaces for {6G} systems: Principles, applications, and research
  directions,'' \emph{IEEE Communications Magazine}, vol.~59, no.~6, pp.
  14--20, 2021.

\bibitem{xianghaoyu2019}
X.~{Yu}, D.~{Xu}, and R.~{Schober}, ``{MISO} wireless communication systems via
  intelligent reflecting surfaces : (invited paper),'' in \emph{2019 IEEE/CIC
  International Conference on Communications in China (ICCC)}, 2019, pp.
  735--740.

\bibitem{shuowen}
S.~{Zhang} and R.~{Zhang}, ``Capacity characterization for intelligent
  reflecting surface aided {MIMO} communication,'' \emph{IEEE Journal on
  Selected Areas in Communications}, vol.~38, no.~8, pp. 1823--1838, 2020.

\bibitem{qingqingwutwc}
Q.~{Wu} and R.~{Zhang}, ``Intelligent reflecting surface enhanced wireless
  network via joint active and passive beamforming,'' \emph{IEEE Transactions
  on Wireless Communications}, vol.~18, no.~11, pp. 5394--5409, 2019.

\bibitem{boyadi2020}
B.~{Di}, H.~{Zhang}, L.~{Song}, Y.~{Li}, Z.~{Han}, and H.~V. {Poor}, ``Hybrid
  beamforming for reconfigurable intelligent surface based multi-user
  communications: Achievable rates with limited discrete phase shifts,''
  \emph{IEEE Journal on Selected Areas in Communications}, vol.~38, no.~8, pp.
  1809--1822, 2020.

\bibitem{cunhua202twc}
C.~{Pan}, H.~{Ren}, K.~{Wang}, W.~{Xu}, M.~{Elkashlan}, A.~{Nallanathan}, and
  L.~{Hanzo}, ``Multicell {MIMO} communications relying on intelligent
  reflecting surfaces,'' \emph{IEEE Transactions on Wireless Communications},
  vol.~19, no.~8, pp. 5218--5233, 2020.

\bibitem{pan2020jsac}
C.~{Pan}, H.~{Ren}, K.~{Wang}, M.~{Elkashlan}, A.~{Nallanathan}, J.~{Wang}, and
  L.~{Hanzo}, ``Intelligent reflecting surface aided {MIMO} broadcasting for
  simultaneous wireless information and power transfer,'' \emph{IEEE Journal on
  Selected Areas in Communications}, vol.~38, no.~8, pp. 1719--1734, 2020.

\bibitem{zhengchu2020}
Z.~{Chu}, W.~{Hao}, P.~{Xiao}, and J.~{Shi}, ``Intelligent reflecting surface
  aided multi-antenna secure transmission,'' \emph{IEEE Wireless Communications
  Letters}, vol.~9, no.~1, pp. 108--112, 2020.

\bibitem{huiming2020}
L.~{Dong} and H.~{Wang}, ``Secure {MIMO} transmission via intelligent
  reflecting surface,'' \emph{IEEE Wireless Communications Letters}, vol.~9,
  no.~6, pp. 787--790, 2020.

\bibitem{hong2020perfectcsi}
S.~{Hong}, C.~{Pan}, H.~{Ren}, K.~{Wang}, and A.~{Nallanathan},
  ``Artificial-noise-aided secure {MIMO} wireless communications via
  intelligent reflecting surface,'' \emph{IEEE Transactions on Communications},
  vol.~68, no.~12, pp. 7851--7866, 2020.

\bibitem{tong2020}
T.~{Bai}, C.~{Pan}, Y.~{Deng}, M.~{Elkashlan}, A.~{Nallanathan}, and
  L.~{Hanzo}, ``Latency minimization for intelligent reflecting surface aided
  mobile edge computing,'' \emph{IEEE Journal on Selected Areas in
  Communications}, vol.~38, no.~11, pp. 2666--2682, 2020.

\bibitem{guizhouwcl}
G.~{Zhou}, C.~{Pan}, H.~{Ren}, K.~{Wang}, M.~D. {Renzo}, and A.~{Nallanathan},
  ``Robust beamforming design for intelligent reflecting surface aided {MISO}
  communication systems,'' \emph{IEEE Wireless Communications Letters}, vol.~9,
  no.~10, pp. 1658--1662, 2020.

\bibitem{xianghaoyu2020}
X.~{Yu}, D.~{Xu}, Y.~{Sun}, D.~W.~K. {Ng}, and R.~{Schober}, ``Robust and
  secure wireless communications via intelligent reflecting surfaces,''
  \emph{IEEE Journal on Selected Areas in Communications}, vol.~38, no.~11, pp.
  2637--2652, 2020.

\bibitem{guirobusttsp}
G.~{Zhou}, C.~{Pan}, H.~{Ren}, K.~{Wang}, and A.~{Nallanathan}, ``A framework
  of robust transmission design for {IRS}-aided {MISO} communications with
  imperfect cascaded channels,'' \emph{IEEE Transactions on Signal Processing},
  vol.~68, pp. 5092--5106, 2020.

\bibitem{hong2020robust}
S.~{Hong}, C.~{Pan}, H.~{Ren}, K.~{Wang}, K.~K. {Chai}, and A.~{Nallanathan},
  ``Robust transmission design for intelligent reflecting surface aided secure
  communication systems with imperfect cascaded {CSI},'' \emph{IEEE
  Transactions on Wireless Communications}, pp. 1--1, 2020.

\bibitem{yuhan2019}
Y.~{Han}, W.~{Tang}, S.~{Jin}, C.~{Wen}, and X.~{Ma}, ``Large intelligent
  surface-assisted wireless communication exploiting statistical {CSI},''
  \emph{IEEE Transactions on Vehicular Technology}, vol.~68, no.~8, pp.
  8238--8242, 2019.

\bibitem{zhangjie}
Z.~Peng, T.~Li, C.~Pan, H.~Ren, W.~Xu, and M.~D. Renzo, ``Analysis and
  optimization for {RIS}-aided multi-pair communications relying on statistical
  {CSI},'' \emph{IEEE Transactions on Vehicular Technology}, vol.~70, no.~4,
  pp. 3897--3901, 2021.

\bibitem{kangda}
K.~Zhi, C.~Pan, H.~Ren, and K.~Wang, ``Statistical {CSI}-based design for
  reconfigurable intelligent surface-aided massive {MIMO} systems with direct
  links,'' \emph{IEEE Wireless Communications Letters}, vol.~10, no.~5, pp.
  1128--1132, 2021.

\bibitem{liyou2021}
L.~You, J.~Xiong, D.~W.~K. Ng, C.~Yuen, W.~Wang, and X.~Gao, ``Energy
  efficiency and spectral efficiency tradeoff in {RIS}-aided multiuser {MIMO}
  uplink transmission,'' \emph{IEEE Transactions on Signal Processing},
  vol.~69, pp. 1407--1421, 2021.

\bibitem{emil2020}
E.~{Bjornson}, O.~{Ozdogan}, and E.~G. {Larsson}, ``Intelligent reflecting
  surface versus decode-and-forward: How large surfaces are needed to beat
  relaying?'' \emph{IEEE Wireless Communications Letters}, vol.~9, no.~2, pp.
  244--248, 2020.

\bibitem{Boulogeorgos2020}
A.~A. {Boulogeorgos} and A.~{Alexiou}, ``Performance analysis of reconfigurable
  intelligent surface-assisted wireless systems and comparison with relaying,''
  \emph{IEEE Access}, vol.~8, pp. 94\,463--94\,483, 2020.

\bibitem{macroojcs}
M.~{Di Renzo}, K.~{Ntontin}, J.~{Song}, F.~H. {Danufane}, X.~{Qian},
  F.~{Lazarakis}, J.~{De Rosny}, D.~T. {Phan-Huy}, O.~{Simeone}, R.~{Zhang},
  M.~{Debbah}, G.~{Lerosey}, M.~{Fink}, S.~{Tretyakov}, and S.~{Shamai},
  ``Reconfigurable intelligent surfaces vs. relaying: Differences,
  similarities, and performance comparison,'' \emph{IEEE Open Journal of the
  Communications Society}, vol.~1, pp. 798--807, 2020.

\bibitem{liangyang2020}
L.~{Yang}, Y.~{Yang}, M.~O. {Hasna}, and M.~S. {Alouini}, ``Coverage,
  probability of {SNR} gain, and {DOR} analysis of {RIS}-aided communication
  systems,'' \emph{IEEE Wireless Communications Letters}, vol.~9, no.~8, pp.
  1268--1272, 2020.

\bibitem{liang2020}
L.~Yang, F.~Meng, J.~Zhang, M.~O. Hasna, and M.~D. Renzo, ``On the performance
  of ris-assisted dual-hop uav communication systems,'' \emph{IEEE Transactions
  on Vehicular Technology}, vol.~69, no.~9, pp. 10\,385--10\,390, 2020.

\bibitem{qintao2020}
Q.~{Tao}, J.~{Wang}, and C.~{Zhong}, ``Performance analysis of intelligent
  reflecting surface aided communication systems,'' \emph{IEEE Communications
  Letters}, vol.~24, no.~11, pp. 2464--2468, 2020.

\bibitem{Ibrahim2021}
H.~{Ibrahim}, H.~{Tabassum}, and U.~T. {Nguyen}, ``Exact coverage analysis of
  intelligent reflecting surfaces with {Nakagami}-m channels,'' \emph{IEEE
  Transactions on Vehicular Technology}, pp. 1--1, 2021.

\bibitem{Atapattu2020}
S.~{Atapattu}, R.~{Fan}, P.~{Dharmawansa}, G.~{Wang}, J.~{Evans}, and T.~A.
  {Tsiftsis}, ``Reconfigurable intelligent surface assisted two-way
  communications: Performance analysis and optimization,'' \emph{IEEE
  Transactions on Communications}, vol.~68, no.~10, pp. 6552--6567, 2020.

\bibitem{jiayi2021}
H.~{Du}, J.~{Zhang}, J.~{Cheng}, and B.~{Ai}, ``Millimeter wave communications
  with reconfigurable intelligent surfaces: Performance analysis and
  optimization,'' \emph{IEEE Transactions on Communications}, pp. 1--1, 2021.

\bibitem{Badiu2020}
M.~{Badiu} and J.~P. {Coon}, ``Communication through a large reflecting surface
  with phase errors,'' \emph{IEEE Wireless Communications Letters}, vol.~9,
  no.~2, pp. 184--188, 2020.

\bibitem{Javier2020}
J.~D.~V. {Sanchez}, P.~{Ramirez-Espinosa}, and F.~J. {Lopez-Martinez},
  ``Physical layer security of large reflecting surface aided communications
  with phase errors,'' \emph{IEEE Wireless Communications Letters}, pp. 1--1,
  2020.

\bibitem{pengxu2020}
P.~{Xu}, G.~{Chen}, G.~{Pan}, and M.~{Di Renzo}, ``Ergodic secrecy rate of
  {RIS}-assisted communication systems in the presence of discrete phase shifts
  and multiple eavesdroppers,'' \emph{IEEE Wireless Communications Letters},
  pp. 1--1, 2020.

\bibitem{dongli2020}
D.~{Li}, ``Ergodic capacity of intelligent reflecting surface-assisted
  communication systems with phase errors,'' \emph{IEEE Communications
  Letters}, vol.~24, no.~8, pp. 1646--1650, 2020.

\bibitem{Durisi2016}
G.~{Durisi}, T.~{Koch}, and P.~{Popovski}, ``Toward massive, ultrareliable, and
  low-latency wireless communication with short packets,'' \emph{Proceedings of
  the IEEE}, vol. 104, no.~9, pp. 1711--1726, 2016.

\bibitem{schiessl2015delay}
S.~Schiessl, J.~Gross, and H.~Al-Zubaidy, ``Delay analysis for wireless fading
  channels with finite blocklength channel coding,'' in \emph{Proceedings of
  the 18th ACM International Conference on Modeling, Analysis and Simulation of
  Wireless and Mobile Systems}, 2015, pp. 13--22.

\bibitem{polyanskiy2010channel}
Y.~Polyanskiy, H.~V. Poor, and S.~Verd{\'u}, ``Channel coding rate in the
  finite blocklength regime,'' \emph{IEEE Transactions on Information Theory},
  vol.~56, no.~5, pp. 2307--2359, 2010.

\bibitem{xiaofang2018}
X.~{Sun}, S.~{Yan}, N.~{Yang}, Z.~{Ding}, C.~{Shen}, and Z.~{Zhong},
  ``Short-packet downlink transmission with non-orthogonal multiple access,''
  \emph{IEEE Transactions on Wireless Communications}, vol.~17, no.~7, pp.
  4550--4564, 2018.

\bibitem{Ghanem2019}
W.~R. {Ghanem}, V.~{Jamali}, Y.~{Sun}, and R.~{Schober}, ``Resource allocation
  for multi-user downlink {URLLC-OFDMA} systems,'' in \emph{2019 IEEE
  International Conference on Communications Workshops (ICC Workshops)}, 2019,
  pp. 1--6.

\bibitem{SHe2108tcom}
C.~{She}, C.~{Yang}, and T.~Q.~S. {Quek}, ``Joint uplink and downlink resource
  configuration for ultra-reliable and low-latency communications,'' \emph{IEEE
  Transactions on Communications}, vol.~66, no.~5, pp. 2266--2280, 2018.

\bibitem{Avranas2018}
A.~{Avranas}, M.~{Kountouris}, and P.~{Ciblat}, ``Energy-latency tradeoff in
  ultra-reliable low-latency communication with retransmissions,'' \emph{IEEE
  Journal on Selected Areas in Communications}, vol.~36, no.~11, pp.
  2475--2485, 2018.

\bibitem{jiechen2019}
J.~{Chen}, L.~{Zhang}, Y.~{Liang}, X.~{Kang}, and R.~{Zhang}, ``Resource
  allocation for wireless-powered {IoT} networks with short packet
  communication,'' \emph{IEEE Transactions on Wireless Communications},
  vol.~18, no.~2, pp. 1447--1461, 2019.

\bibitem{yulinhu2019}
Y.~{Hu}, Y.~{Zhu}, M.~C. {Gursoy}, and A.~{Schmeink}, ``{SWIPT}-enabled
  relaying in {IoT} networks operating with finite blocklength codes,''
  \emph{IEEE Journal on Selected Areas in Communications}, vol.~37, no.~1, pp.
  74--88, 2019.

\bibitem{hongrentwc}
H.~{Ren}, C.~{Pan}, Y.~{Deng}, M.~{Elkashlan}, and A.~{Nallanathan}, ``Joint
  power and blocklength optimization for {URLLC} in a factory automation
  scenario,'' \emph{IEEE Transactions on Wireless Communications}, vol.~19,
  no.~3, pp. 1786--1801, 2020.

\bibitem{hongrenjsac}
------, ``Joint pilot and payload power allocation for massive-{MIMO}-enabled
  {URLLC} {IIoT} networks,'' \emph{IEEE Journal on Selected Areas in
  Communications}, vol.~38, no.~5, pp. 816--830, 2020.

\bibitem{Zheng2019}
J.~{Zheng}, Q.~{Zhang}, and J.~{Qin}, ``Average block error rate of downlink
  {NOMA} short-packet communication systems in nakagami- $m$ fading channels,''
  \emph{IEEE Communications Letters}, vol.~23, no.~10, pp. 1712--1716, 2019.

\bibitem{Schiessl2018}
S.~{Schiessl}, H.~{Al-Zubaidy}, M.~{Skoglund}, and J.~{Gross}, ``Delay
  performance of wireless communications with imperfect {CSI} and finite-length
  coding,'' \emph{IEEE Transactions on Communications}, vol.~66, no.~12, pp.
  6527--6541, 2018.

\bibitem{jiezeng2020}
J.~{Zeng}, T.~{Lv}, R.~P. {Liu}, X.~{Su}, Y.~J. {Guo}, and N.~C. {Beaulieu},
  ``Enabling ultrareliable and low-latency communications under shadow fading
  by massive {MU-MIMO},'' \emph{IEEE Internet of Things Journal}, vol.~7,
  no.~1, pp. 234--246, 2020.

\bibitem{hongren2019wcl}
H.~{Ren}, C.~{Pan}, K.~{Wang}, Y.~{Deng}, M.~{Elkashlan}, and A.~{Nallanathan},
  ``Achievable data rate for {URLLC}-enabled {UAV} systems with {3-D} channel
  model,'' \emph{IEEE Wireless Communications Letters}, vol.~8, no.~6, pp.
  1587--1590, 2019.

\bibitem{shuhan2019}
S.~{Han}, X.~{Xu}, Z.~{Liu}, P.~{Xiao}, K.~{Moessner}, X.~{Tao}, and
  P.~{Zhang}, ``Energy-efficient short packet communications for uplink
  {NOMA}-based massive {MTC} networks,'' \emph{IEEE Transactions on Vehicular
  Technology}, vol.~68, no.~12, pp. 12\,066--12\,078, 2019.

\bibitem{Ranjha2020}
A.~{Ranjha} and G.~{Kaddoum}, ``{URLLC} facilitated by mobile {UAV} relay and
  {RIS}: A joint design of passive beamforming, blocklength and {UAV}
  positioning,'' \emph{IEEE Internet of Things Journal}, pp. 1--1, 2020.

\bibitem{Walid2020}
\BIBentryALTinterwordspacing
W.~R. Ghanem, V.~Jamali, and R.~Schober, ``Joint beamforming and phase shift
  optimization for multicell {IRS}-aided {OFDMA-URLLC} systems.'' [Online].
  Available: \url{https://arxiv.org/abs/2010.07698}
\BIBentrySTDinterwordspacing

\bibitem{shannon2001mathematical}
C.~Shannon, ``A mathematical theory of communication,'' \emph{The Bell System
  Technical Journal}, vol.~27, no.~1, pp. 379--423, July 1948.

\bibitem{HypergeometricPFQ}
\BIBentryALTinterwordspacing
Wolfram, ``Generalized hypergeometric function.'' [Online]. Available:
  \url{{https://reference.wolfram.com/language/ref/HypergeometricPFQ.html}}
\BIBentrySTDinterwordspacing

\bibitem{Poly}
\BIBentryALTinterwordspacing
Wikipedia, ``Polygamma approximation.'' [Online]. Available:
  \url{{{https://en.wikipedia.org/wiki/Polygamma\_function}}}
\BIBentrySTDinterwordspacing

\bibitem{gradshteyn2014table}
I.~S. Gradshteyn and I.~M. Ryzhik, \emph{Table of integrals, series, and
  products}.\hskip 1em plus 0.5em minus 0.4em\relax Academic press, 2014.

\bibitem{ExpIntegralE}
\BIBentryALTinterwordspacing
Wolfram, ``Exponential integral function.'' [Online]. Available:
  \url{{https://reference.wolfram.com/language/ref/ExpIntegralE.html?q=ExpIntegralE}}
\BIBentrySTDinterwordspacing

\bibitem{Ricianfunchde}
\BIBentryALTinterwordspacing
``Rice distribution.'' [Online]. Available: \url{https://en.wikipedia.org/wiki}
\BIBentrySTDinterwordspacing

\bibitem{Emilwcl2021}
E.~Bjornson and L.~Sanguinetti, ``Rayleigh fading modeling and channel
  hardening for reconfigurable intelligent surfaces,'' \emph{IEEE Wireless
  Communications Letters}, vol.~10, no.~4, pp. 830--834, 2021.

\bibitem{tulino2004random}
A.~M. Tulino, S.~Verd{\'u}, and S.~Verdu, \emph{Random matrix theory and
  wireless communications}.\hskip 1em plus 0.5em minus 0.4em\relax Now
  Publishers Inc, 2004.

\end{thebibliography}


\end{document}